\documentclass[<preprint>,sort&compress,numbers]{elsarticle} 

\usepackage{graphicx}
\usepackage{amsmath}
\usepackage[version=4]{mhchem}
\usepackage{siunitx}
\usepackage{longtable,tabularx}

\usepackage{bm}
\usepackage{mathrsfs}
\usepackage{epsfig}
\usepackage{subfigure}
\usepackage{setspace}
\usepackage{cancel} 
\usepackage{appendix}
\usepackage{color}
\usepackage{wrapfig}
\usepackage{xspace}

\usepackage{amsfonts}		
\usepackage{rotating}     

\usepackage{caption}
\usepackage{arydshln}
\usepackage{indentfirst}
\usepackage{float}
\usepackage{url}
\usepackage{array}
\usepackage{times}
\usepackage{booktabs} 

\newcommand{\ignore}[1]{}

\newcommand{\ie}[0]{{i.e.\@}\xspace}

\newcommand{\bbm}{\begin{bmatrix}}
\newcommand{\ebm}{\end{bmatrix}}
\newcommand{\bma}[1]{\left[\begin{array}{#1}}
\newcommand{\ema}{\end{array}\right]}
\newcommand{\vect}{\underrightarrow}

\DeclareMathAlphabet{\mbf}{OT1}{ptm}{b}{n}
\newcommand{\mbs}[1]{{\boldsymbol{#1}}}

\newcommand{\mbsbar}[1]{{\bar{\boldsymbol{#1}}}}

\newcommand{\mbstilde}[1]{{\tilde{\boldsymbol{#1}}}}
\newcommand{\mbfbar}[1]{{\bar{\mbf{#1}}}}

\newcommand{\mbftilde}[1]{{\tilde{\mbf{#1}}}}

\def\fdota{{\raisebox{-2pt}{\LARGE $\cdot$}}}
\def\fdotb{{\raisebox{-0.6ex}{ \kern0.2ex\raisebox{0.8ex}{\tiny $\hspace*{-1ex}\circ$}}}}

\def\fddotb{{\raisebox{-0.6ex}{ \kern0.2ex\raisebox{0.8ex}{\tiny $\hspace*{-1ex}\circ\circ$}}}}
\newcommand{\fdot}[1]{{^{\fdota{\mbox{\scriptsize${#1}$}}}}}

\newcommand{\p}{\partial}
\newcommand{\f}{\frac}

\newcommand{\ura}[1]{{\underrightarrow{{#1}}}}
 
\newcommand{\dee}{\textrm{d}}

\newcommand{\trans}{{\ensuremath{\top}}} 
\newcommand{\beq}{\begin{equation}}
\newcommand{\eeq}{\end{equation}}
\newcommand{\bdis}{\begin{displaymath}}
\newcommand{\edis}{\end{displaymath}}
\newcommand{\beqarray}{\begin{eqnarray}}
\newcommand{\eeqarray}{\end{eqnarray}}
\newcommand{\beqarraynn}{\begin{eqnarray*}}
\newcommand{\eeqarraynn}{\end{eqnarray*}}

\usepackage{listings}
\usepackage{comment}
\DeclareMathOperator*{\minimize}{minimize}

\journal{Acta Astronautica} 
\begin{document}

\begin{frontmatter}
\title{Solar Sail Momentum Management With Mass Translation and Reflectivity Devices Using Predictive Control}

\author[]{Ping-Yen Shen}
\ead{shen0251@umn.edu}
\author[]{Ryan J. Caverly\corref{cor1}}
\ead{rcaverly@umn.edu}

\cortext[cor1]{Corresponding author}

\address{Department of Aerospace Engineering and Mechanics, University of Minnesota, 110 Union St. SE, Minneapolis, MN 55455, United States}

\begin{abstract}
Solar sails enable propellant-free space missions by utilizing solar radiation pressure as thrust. However, disturbance torques act on the solar sail and effective attitude control leads to the continuous accumulation of reaction wheel angular momentum, necessitating an efficient momentum management strategy to prevent saturation.
This paper presents a novel momentum management controller using model predictive control (MPC) that is tailored for solar sails, accommodating the unique actuation mechanisms of an active mass translator (AMT) and reflectivity control devices (RCDs). A first-order hold discretization and tailored motion costs are applied to the AMT translation, while the RCD actuation is handled using pulse-width modulation (PWM)-inspired quantization to address their on-off inputs. To enhance prediction accuracy, an iterative backwards-in-time MPC approach is introduced, incorporating the effects of PWM-quantized inputs into the optimization process.
The dynamic model accounts for the time-dependent center of mass and moment of inertia changes caused by AMT translation, extending its applicability to other spacecraft with mass-shifting actuators. Simulation results demonstrate the effectiveness of the proposed framework in reaction wheel desaturation, attitude control, and momentum management actuation efficiency, highlighting the potential of integrating MPC to manage coupled nonlinear dynamics and discrete actuator constraints for solar sails.
\end{abstract}


\end{frontmatter}

\section{Introduction}
Space sailing is a field of growing interest with diverse applications in space science and engineering~\cite{berthet2024space}.
Considering the growing number of space debris~\cite{zhang2023long, bianchi2024preliminary}, drag sails have been proposed for the deorbiting end-of-life satellites~\cite{bigdeli2025mechanics, guglielmo2019drag, nagavarapu2025cubesats}.
On the other hand, the concept of a solar sail introduces the opportunity for fuel-efficient, and potentially even fuel-free, inter-planetary travel and deep-space exploration~\cite{eldad2015propellantless,farres2023propellant,miller2022high}.
The momentum transferred to a solar sail by solar radiation pressure (SRP) provides a promising source of propulsive thrust that can be controlled by adjusting the attitude of the solar sail, and thus the direction and magnitude of thrust. 
A larger area-to-mass ratio increases a solar sail's efficiency, which motivates the desire of developing a large-scale solar sail system, such as NASA's Solar Cruiser, which involves a sail area greater than $1600$~m$^2$~\cite{JohnsonLes2020SCTM,johnson2019solar,pezent2021preliminary,johnson2022nasa}.
However, larger-scale lightweight solar sails are inherently flexible, resulting in undesirable structural deformations that shift the solar sail's center of pressure (CP) and create disturbance torques that need to be rejected by an attitude control system~\cite{gauvain2023solar,fu2015attitude,firuzi2018attitude,wie2004solar1,wie2004solar2,ingrassia2013solar}. Reaction wheels (RWs) are typically used for this purpose, but eventually suffer from an accumulation of stored angular momentum, which causes a loss of attitude control authority when the RWs saturate, necessitating momentum management. Standard momentum management strategies for spacecraft, such as the use of thrusters or magnetic actuation, are not well-suited for solar sails, since they either require onboard fuel or the presence of an external magnetic field. To truly take advantage of the potential of solar sails for fuel-free deep-space exploration, more creative momentum management solutions are required.

To solve this challenge, Solar Cruiser uses an active mass translator (AMT) and reflectivity control devices (RCDs) as actuators to respectively generate pitch/yaw and roll torques for momentum management~\cite{Inness2023,inness2024controls}. The AMT serves as a mechanism between two portions of the spacecraft bus, allowing them to translate relative to each other in a plane. This results in a controllable shift of the solar sail's center of mass (CM), allowing for the generation of SRP-induced torques or the ``trimming'' out of disturbance torques in the pitch and yaw axes. Unlike rotative tip-devices~\cite{guerrant2015tactics,choi2016control,abrishami2020optimized,hassanpour2020collocated}, the RCDs are stationary thin film membranes located towards the extremities of the solar sail membrane and are canted at an angle to the plane of the sail membrane. They generate roll torques about the axis normal to the sail by adjusting their reflectivity when a voltage is applied to them. A challenge in the operation of RCDs is that they operate in an on-off fashion, either generating no torque or a constant magnitude torque in the positive or negative roll direction. 
Solar Cruiser's attitude determination and control system  
includes control subsystems for momentum management that command the AMT and RCDs~\cite{Tyler2024}. The current design of this momentum management system involves decoupled proportional-integral-derivative (PID) controllers to actuate the two axes of the AMT based on the momentum stored in the pitch/yaw axis RWs, as well as an on-off actuation strategy for the RCDs~\cite{Inness2023,Tyler2024}. 
Although this simple control strategy was shown to successfully perform momentum management, it neglects the dynamic coupling between the AMT and RCD actuators, which can result in undesirable performance bordering on closed-loop instability~\cite{Inness2023}. Additionally, the on-off nature of the RCDs is handled in the approach from~\cite{Inness2023,Tyler2024} using activation/deactivation thresholds, which creates additional nonlinearities in the feedback system, affecting closed-loop performance and stability properties. There is a pressing need for a momentum management control policy that is capable of explicitly accounting for the nonlinear, coupled dynamics involved in the AMT and RCD actuators, as well as the practical constraints associated with their operation.

Model predictive control (MPC)~\cite{camacho2007constrained,rawlings2017model} has been used extensively in industry and academia~\cite{morari1999model,mayne2014model}.
It is typically implemented by solving an online optimization problem that involves constraints based on the system dynamics, as well as limits on the allowable states and control inputs. 
With the continuous improvement of computation capability, MPC has become an option for onboard real-time control in modern aerospace applications~\cite{eren2017model,di2018real,petersen2023safe,lee2017geometric}, although oftentimes the MPC formulation has to be considered carefully to ensure real-time capabilities.  
For example,~\cite{MAMMARELLA2018585} presented a framework using an off-line MPC policy for trajectory generation and an online MPC policy for robust control. Another example involves the use of nonlinear inner-loop attitude feedback controller for a geostationary Earth orbit satellite with an outer-loop MPC policy running at a slower rate for combined station keeping and momentum management~\cite{caverly2020electric}. 
An example of the capability provided by MPC for spacecraft applications includes the use of expected environmental disturbances as a means of actuation, such as actuating solar panels to interact with SRP torques to control an underactuated spacecraft~\cite{jin2022model}, using the limited actuation provided by magnetic torque rod for constrained attitude control~\cite{Halverson2024}, and leveraging atmospheric drag modulation to control the trajectory of a spacecraft in low-Earth orbit~\cite{hayes2023model}.

In this work, MPC is proposed for solar sail momentum management as a means to account for the environmental disturbance torques acting on the solar sail, as well as the coupled nature of the actuators and the nonlinearities of the system's dynamics.
Given the limited flight hardware and computational resources onboard, a simple dynamic model and convex formulation are necessary for real-time implementation.
The unique actuation properties of the AMT and RCDs make this MPC formulation challenging, resulting in an optimization problem that is naturally non-convex and unsuitable for implementation on flight hardware.
To remedy this issue, we first derive a dynamic model of an AMT equipped solar sail, accounting for the effect of changing the solar sail's CM. This provides an accurate prediction model that can be used within an MPC policy. The motion of the AMT is then discretized with a first-order-hold (FOH) allowing for piece-wise linear motion of the AMT to be accurately captured in the prediction model. 
The RCDs operate in an on-off fashion, which traditionally would manifest as integer variables within an MPC optimization scheme. 
This type of optimization problem is highly non-convex and is usually solved using dynamic programming, mixed-integer programming, or branch-and-bound approach~\cite{martins2021engineering,axehill2006mixed,axehill2010improved,botelho2024explicit}, all of which demand significant computational memory and time. To avoid this impractical approach, we propose solving an initial MPC problem where the RCD torque is allowed to vary continuously between its positive and negative actuation magnitude, then quantizing the result into a single on-off pulse. This initial problem can be solved as a quadratic problem (QP) in an efficient fashion using a linear approximation of the system's dynamics. A pulse-width-modulation (PWM) quantization inspired by the thruster momentum management actuation in~\cite{chegeni2014attitude,zlotnik2017mpc,caverly2018off,caverly2020electric,botelho2024explicit} is then applied to turn the optimal continuous RCD input solution from MPC into a single discrete pulse with varying pulse length (duration) within a single timestep. The result is an MPC policy that can be computed efficiently and reliable, while ensuring a physically-realizable quantized RCD input.

A second version of the proposed MPC-based momentum management strategy is presented in this work to address the fact that performing PWM quantization on the MPC solution may result sub-optimal performance. This is addressed through a novel iterative backwards-in-time MPC strategy that fixes the last RCD input in the prediction horizon as a quantized value and re-optimizes the remainder of the control inputs. This process is repeated by fixing the next-to-last RCD input as a quantized value, and continuing this process until eventually the first RCD input is quantized. 
This iterative approach removes the complexity of a branch-and-bound approach to integer programming, while including the quantization knowledge in the MPC prediction model to obtain improved performance. 

The contributions of this paper are highlighted as 1) the dynamic modeling of a solar sail equipped with AMT accounting for the effects of a moving CM; 2) two solar sail momentum management strategies using MPC that meet practical operational needs through the explicit inclusion of actuator magnitude, actuator rate limits, and quantization of the RCD inputs, which is shown to significantly reduce the actuation required compared to the state-of-the-art method in~\cite{Inness2023,Tyler2024}; and 3) a novel backwards-in-time iterative solution approach to the second MPC strategy that incorporates knowledge of the quantized RCD input into the MPC optimization problem in a computationally-efficient manner by solving a sequence of convex optimization problems. 
Together, these contributions represent an advance in state-of-the-art solar sail AMT/RCD momentum management, which up until now has struggled to meet operational requirements due to the use of decoupled PID control loops with on-off thresholding.
The practical challenges associated with the momentum management actuators and a detailed derivation of the dynamic model are presented in Section~\ref{sec:sys_dynamics}.
In Section~\ref{sec:MPC}, an MPC-based momentum management controller is formulated to meet the operational requirements of the solar sail and two different methods are proposed to address the quantized RCD inputs. 
Numerical simulations and comparisons between the proposed methods using practical solar sail parameters are presented in Section~\ref{sec:Num_Sim}. 


\section{Notation} \label{sec:notation}

In this paper, reference frame $\mathcal{F}_a$ is defined by three orthonormal, dextral basis vectors $\vect{a}^1$, $\vect{a}^2$, $\vect{a}^3$. The position vector $\vect{r}^{ij}$ describes the position of point $i$ relative to point $j$ and its components when resolved in reference frame $\mathcal{F}_a$ are denoted by $\mbf{r}^{ij}_a = \bbm r^{ij}_{a1} & r^{ij}_{a2} & r^{ij}_{a3} \ebm^\trans$. The cross product $\vect{w} = \vect{u} \times \vect{v}$ resolved in $\mathcal{F}_a$ is expressed as $\mbf{u}^\times_a \mbf{v}_a$, where
\beq
\mbf{u}^\times_a = \bbm u_{a1} \\ u_{a2} \\ u_{a3} \ebm^\times = \bbm 0 & -u_{a3} & u_{a2}\\u_{a3} & 0 & -u_{a1}\\-u_{a2} & u_{a1} & 0 \ebm
\eeq
is the skew-symmetric cross-product matrix of $\mbf{u}_a$. The direction cosine matrix (DCM) $\mbf{C}_{ba}$ describes the orientation of reference frame $\mathcal{F}_b$ relative to $\mathcal{F}_a$ and can be used to convert a vector resolved in one frame to another (e.g., $\mbf{u}_b = \mbf{C}_{ba} \mbf{u}_a$).

\section{System Dynamics} \label{sec:sys_dynamics}

This section presents the nonlinear dynamics of a solar sail equipped with three RWs as attitude control actuators, and with AMT and RCDs as momentum management actuators. 
A solar sail body-fixed frame $\mathcal{F}_{b}$ is defined, where $\vect{b}^3$ is normal to the sail membrane, while $\vect{b}^2$ and $\vect{b}^1$ denote the in-plane pitch and yaw axes, respectively. 
The RW system is designed to nominally perform attitude control using a PID control law, operating at a relatively high frequency ($1$~Hz in the numerical simulations of Section~\ref{sec:Num_Sim}). In contrast, the momentum management system operates at a lower frequency ($0.01$~Hz in the numerical simulations of Section~\ref{sec:Num_Sim}), focusing primarily on unloading the angular momentum accumulated in the RWs due to external disturbances. 

Before deriving the detailed solar sail dynamics, Section~\ref{Sec:momentum management_actuators} provides an overview of the momentum management actuators utilized in NASA’s Solar Cruiser: the AMT and RCDs. A derivation of the solar sail's attitude dynamics, accounting for the effects of AMT translation on the spacecraft's CM and moment of inertia, is presented in Sections~\ref{Sec:nonlin_dyn} and~\ref{sec:RW_PID}. A linearization procedure for the derived nonlinear model is detailed in Section~\ref{sec:Linearization}, serving as the prediction model for MPC. The nonlinear model itself is subsequently used in the numerical simulations of Section~\ref{sec:Num_Sim}.
Unlike conventional rigid-body spacecraft dynamics, the solar sail introduces unique challenges. The AMT’s ability to translate a significant portion of the spacecraft’s mass creates dynamics that are directly dependent on the AMT’s position. This dependency is highlighted throughout the modeling and control synthesis processes in Sections~\ref{Sec:nonlin_dyn} and~\ref{sec:MPC}.

\subsection{Momentum Management Actuators} \label{Sec:momentum management_actuators}

The focus of this paper is to design a momentum management strategy that efficiently unloads the angular momentum accumulated in the RWs. To achieve this, AMT and RCDs are employed as the primary momentum management actuators on NASA’s Solar Cruiser~\cite{Inness2023}, which are scaled to perform control authority $> 1.5$ times of the disturbance torque~\cite{johnson2022nasa}.

\begin{figure}[h!]
    \centering
        \includegraphics[width=0.75\textwidth]{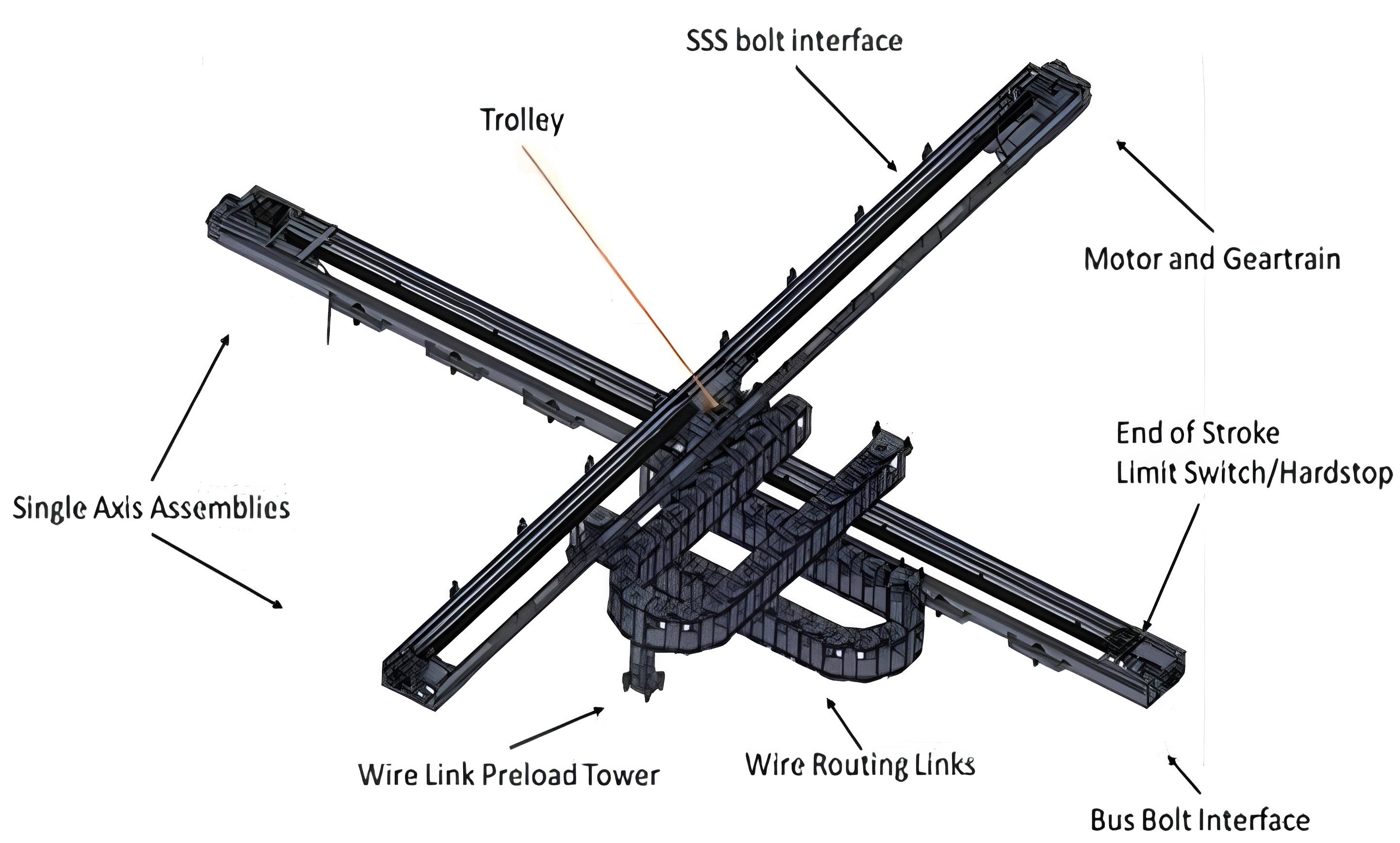}
    \vspace{-8pt}
    \caption{Depiction of the active mass translator (AMT). Image Credit: NASA~\cite{inness2024controls}}\label{Fig:AMT}
\end{figure}

An active mass translator (AMT), depicted in Figure~\ref{Fig:AMT}, moves a portion of the spacecraft bus mass within a plane parallel to the sail surface. By adjusting the AMT’s position, the relative positions of the CM and the CP are modified, creating a moment arm when SRP acts on the sail surface. This generates torques about the pitch and yaw axes (respectively denoted as the $\vect{b}^2$ and $\vect{b}^1$ axes in Section~\ref{Sec:nonlin_dyn}). 
Solar Cruiser's AMT translates approximately $50$\% of the spacecraft’s mass, enabling a substantial CM displacement within the AMT travel range of $\pm 0.29$~m~\cite{JohnsonLes2020SCTM}. This torque can mitigate RW angular momentum buildup and counteract disturbances by moving the AMT to the appropriate position. However, the CM displacement caused by AMT motion also alters the spacecraft's moment of inertia and overall dynamics. The derivation of a dynamic model accounting for these effects is detailed in Section~\ref{Sec:nonlin_dyn}. 
In this paper, the AMT’s control input is modeled to reflect realistic motion constraints. Instead of a zeroth-order hold (ZOH), which assumes abrupt positional changes, a FOH is used to simulate continuous translation between momentum management time steps. This representation also facilitates the inclusion of AMT translation rate limits as constraints in the MPC framework.
Details of the discretization and linear interpolation are presented in Section~\ref{sec:Discretization}. 

\begin{figure}[h!]
    \centering
    \subfigure[]
{
        \includegraphics[width=0.75\textwidth]{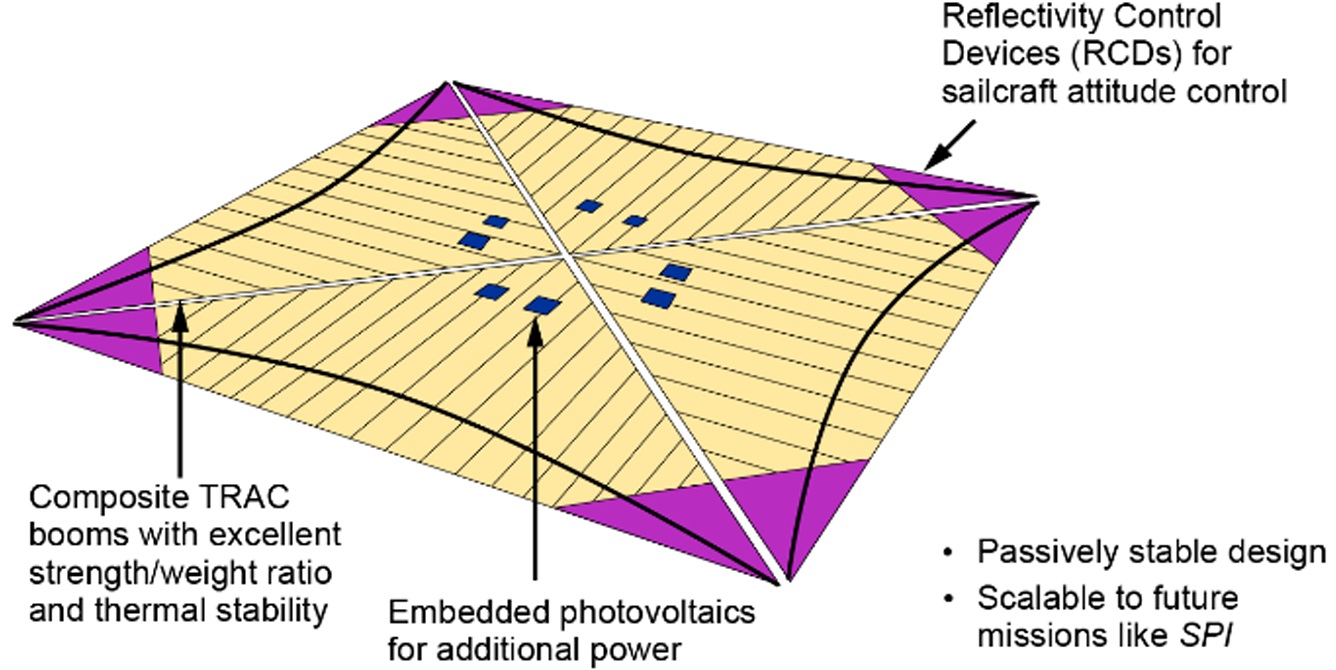}
}
    \subfigure[]
{
        \includegraphics[width=0.75\textwidth]{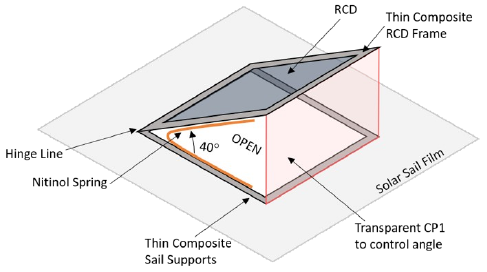}
}
    \vspace{-8pt}
    \caption{(a) A depiction of the reflectivity control devices (RCDs) embedded in a solar sail membrane. (b) A close-up schematic of the tenting setup used for the RCDs. Image Credits: NASA~\cite{johnson2019solar,JohnsonLes2020SCTM}}\label{Fig:RCD}
\end{figure}

Reflectivity control devices (RCDs) serve as the key momentum management actuators for the roll axis (denoted as the $\vect{b}^3$ axis in Section~\ref{Sec:nonlin_dyn}). Each RCD operates by modulating its reflectivity using small electrical power inputs, altering the distribution of SRP on the sail and generating torques for roll-axis control.
A total of eight RCDs are strategically positioned in pairs near the outer edges of each sail boom, as illustrated in Figure~\ref{Fig:RCD}. 
These pairs are arranged in an inclined ``tent-like'' configuration with opposing angles. Turning on four of the RCDs with the same tent angle induces a roll-axis torque due to an imbalance in the differential forces in-plane to the sail. The differential out-of-plane forces induced by each of the RCDs are canceled out due to symmetry. This torque, denoted as $\mbs{\tau}_b^{\text{RCD}}$, is treated as a pure roll-axis control input in the dynamic model described in Section~\ref{Sec:nonlin_dyn}. Disturbances arising from sail surface imperfection, roughness, or deformation are modeled separately as disturbance torque $\mbs{\tau}_b^{\text{dist}}$ in Section~\ref{Sec:nonlin_dyn}. 
The unique actuation mechanism of RCDs requires careful modeling due to their discrete, electrically powered operation. They function in on-off actuation pulses, represented by integer values $\alpha_\text{on-off} \in \{-1, 0, 1\}$ corresponding to negative, zero, and positive directions of a fixed magnitude RCDs torque $\mbs{\tau}_{b,\text{on}}^\text{RCD}$, \ie, $\mbs{\tau}_b^{\text{RCD}} = \alpha_\text{on-off} \cdot \mbs{\tau}_{b,\text{on}}^\text{RCD}$. This discrete nature introduces a non-convex integer constraint, presenting computational challenges for efficient control optimization. Approaches to address these constraints, including PWM-quantization and iterative backtracking methods, are discussed in Sections~\ref{sec:PWM} and~\ref{sec:iter_back}.

The AMT and RCDs constitute the primary momentum management actuators for the solar sail spacecraft. This paper considers the AMT position and the collective torque generated by RCDs as the primary momentum management inputs, forming the foundation for the proposed control framework.

\subsection{Nonlinear System Dynamics} \label{Sec:nonlin_dyn}

Let $\mathcal{F}_{a}$, defined by basis vectors $\vect{a}^1$, $\vect{a}^2$, $\vect{a}^3$, be the inertial reference frame. 
A solar sail body-fixed frame $\mathcal{F}_{b}$ is defined as $\vect{b}^3$ pointing through the roll axis of the sail, and $\vect{b}^2$, $\vect{b}^1$ denote the pitch and yaw axes, respectively. 
A 3-2-1 Euler angle sequence is used to describe the rotation between $\mathcal{F}_{a}$ and $\mathcal{F}_{b}$, so that $\mbf{C}_{ba} = \mbf{C}_1(\theta_1) \mbf{C}_2(\theta_2) \mbf{C}_3(\theta_3)$ is the DCM describing the orientation of $\mathcal{F}_{b}$ relative to $\mathcal{F}_{a}$, and $\mbs{\omega}^{ba}_b = \Big[\omega^{ba}_{b1} \,\,\, \omega^{ba}_{b2} \,\,\, \omega^{ba}_{b3} \Big]^\trans= \mbf{S}(\mbs{\theta})\dot{\mbs{\theta}} $
is the angular velocity of $\mathcal{F}_{b}$ relative to $\mathcal{F}_{a}$ resolved in $\mathcal{F}_{b}$, where $\mbs{\theta} = \Big[ \theta_1 \,\,\, \theta_2 \,\,\, \theta_3 \Big]^\trans$ is the set of yaw, pitch, and roll Euler angles, and $\mbf{S}(\mbs{\theta})$ is the mapping matrix between angular rates $\dot{\mbs{\theta}}$, given by
\bdis
\mbf{S}(\mbs{\theta}) = \bbm 1 & 0 & -\sin(\theta_2) \\ 0 & \cos(\theta_1) & \sin(\theta_1)\cos(\theta_2) \\ 0 & -\sin(\theta_1) & \cos(\theta_1)\cos(\theta_2) \ebm.
\edis
It is worth noting that $\mbf{S}(\mbs{\theta})$ depends only on $\theta_1$ and $\theta_2$ due to the selected 3-2-1 Euler angle sequence. Given that a solar sail is designed to keep the Sun within its field of view and maintain a nominal spin about the $\vect{b}^3$ axis, this choice allows for ease of linearization about any nominal angular velocity about the $\vect{b}^3$ axis, and positions the kinematic singularity at $180\text{\textdegree}$ from the nominal inertial pointing attitude. 

\begin{figure}[h!]
    \centering
        \includegraphics[width=0.75\textwidth]{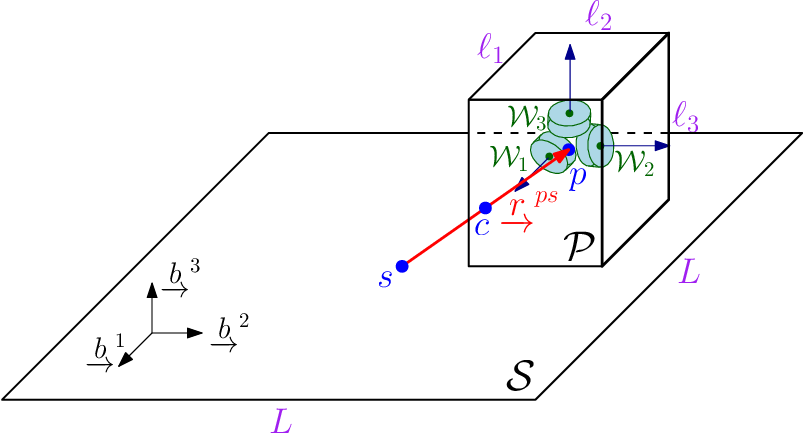}
    \vspace{-8pt}
    \caption{Conceptualized solar sail model with AMT translation between bus $\mathcal{P}$ and sail $\mathcal{S}$ (not drawn to scale), where the position vector is denoted by $\ura{{r}}^{ps}$. }\label{Fig:dynamic_model}
\end{figure}

Consider a rigid body spacecraft $\mathcal{B}$, as shown in Figure~\ref{Fig:dynamic_model}, consisting of a square planar sail membrane $\mathcal{S}$ with mass $m_s$, and a rectangular cuboid bus $\mathcal{P}$ with mass $m_p$.
The dimension of bus $\mathcal{P}$ is given by $\ell_1$, $\ell_2$, $\ell_3$ in each edge. The square sail $\mathcal{S}$ has length $L$ in each edge.
For simplicity, it is assumed that both $\mathcal{S}$ and $\mathcal{P}$ have their CM colocated with their geometric center, denoted by points $s$ and $p$.
The CM of the entire spacecraft $\mathcal{B}$ is denoted by point $c$.

The bus $\mathcal{P}$ houses the attitude control system with three RWs (denoted by $\mathcal{W}_1$, $\mathcal{W}_2$, $\mathcal{W}_3$) located at point $p$, which is coincident with the CM of the bus. Each of the RWs spins about one of the three nominal axes $\vect{b}^1$, $\vect{b}^2$, $\vect{b}^3$ of the spacecraft body frame $\mathcal{F}_{b}$. These assumptions can be relaxed in principle, but are chosen to simplify portions of the system dynamics in this work. 
Point $p$ relative to point $s$ resolved in frame $\mathcal{F}_{b}$ is denoted as $\mbf{r}^{ps}_b$.
This position vector is determined by the AMT's translation of the bus $\mathcal{P}$ relative to the sail $\mathcal{S}$ within the $\ura{b}^1-\ura{b}^2$ plane, such that $\mbf{r}^{ps}_b(t) = \Big[ r^{\text{AMT}}_{b1}(t) \,\,\, r^{\text{AMT}}_{b2}(t) \,\,\, {r}^{ps}_{b3} \Big]^\trans$, where the distance ${r}^{ps}_{b3}$ is constant. 

Since the dimension of the sail $\mathcal{S}$ is much greater than the dimension of the bus $\mathcal{P}$, \ie, $L \gg \{\ell_1, \ell_2, \ell_3\}$, it is reasonable to assume that the RWs are colocated at point $p$.
The moment of inertia of $\mathcal{P}$ and $\mathcal{S}$ relative to each of their own CM (point $p$ and point $s$) respectively are $\mbf{J}^{\mathcal{P}p}_b = \text{diag} \big(\f{m_p}{12}(\ell_2^2+\ell_3^2), \f{m_p}{12}(\ell_1^2+\ell_3^2), \f{m_p}{12}(\ell_1^2+\ell_2^2) \big)$ and $\mbf{J}^{\mathcal{S}s}_b = \text{diag} \big(\f{m_s}{12} (L^2+T^2), \f{m_s}{12} (L^2+T^2), \f{m_s}{6} L^2 \big)$, where $T$ is the thickness of the sail membrane. 
Applying the parallel axis theorem, the moment of inertia of $\mathcal{P}$ and $\mathcal{S}$ relative to point $c$ (the CM of spacecraft $\mathcal{B}$) are respectively 
    \begin{align}
        &\mbf{J}_b^{\mathcal{P}c}(t) = \mbf{J}^{\mathcal{P}p}_b - m_p\mbf{r}_b^{{pc}^\times}(t)\mbf{r}_b^{{pc}^\times}(t) ,
        \\
        &\mbf{J}_b^{\mathcal{S}c}(t) = \mbf{J}^{\mathcal{S}s}_b - m_s\mbf{r}_b^{{sc}^\times}(t)\mbf{r}_b^{{sc}^\times}(t), \label{eq:J}
    \end{align}
where $\mbf{r}_b^{pc}(t) = -\f{m_p}{m_p+m_s}\mbf{r}^{ps}_b(t)$ and $\mbf{r}_b^{sc}(t) = \f{m_s}{m_p+m_s}\mbf{r}^{ps}_b(t)$ depend on the mass ratio between bus $\mathcal{P}$ and sail $\mathcal{S}$, as well as the position of the AMT. 
The moment of inertia of the whole spacecraft $\mathcal{B}$ relative to the collective CM of the spacecraft (point $c$ in Figure~\ref{Fig:dynamic_model}) is
    \beq
        \mbf{J}_b^{\mathcal{B}c}(t) = \mbf{J}_b^{\mathcal{P}c}(t) + \mbf{J}_b^{\mathcal{S}c}(t) = \mbf{J}^{\mathcal{P}p}_b + \mbf{J}^{\mathcal{S}s}_b - \f{m_p^3+m_s^3}{(m_p+m_s)^2}\mbf{r}_b^{{ps}^\times}(t)\mbf{r}_b^{{ps}^\times}(t) .
    \eeq
The bus $\mathcal{P}$ can be split into the platform portion (denoted $\mathcal{G}$) and the RW portion (denoted $\mathcal{W}_i$, $i=1,2,3$), such that the mass of the bus is $m_p = m_g + 3m_w$, where $m_g$ is the mass of the platform and $m_w$ is the mass of each RW. The angular momentum of the bus relative to point $c$ is split into $\vect{h}^{\mathcal{P}c/a}(t) = \vect{h}^{\mathcal{G}c/a}(t) + \sum_{i=1}^3 \vect{h}^{\mathcal{W}_ic/a}(t)$ and the moment of inertia matrix about the bus CM (point $p$) is $\mbf{J}_b^{\mathcal{P}p} = \mbf{J}_b^{\mathcal{G}p} + \sum_{i=1}^3 \mbf{J}^{\mathcal{W}_ip}_b$. 
The angular momentum of the sail $\mathcal{S}$ and the platform $\mathcal{G}$, respectively, about about the spacecraft's CM (point $c$) with respect to inertial reference frame $\mathcal{F}_{a}$ and resolved in $\mathcal{F}_b$ are given by~\cite{schaub2018analytical}
\begin{align}
    \mbf{h}_b^{\mathcal{S}c/a}(t) &= \mbf{J}^{\mathcal{S}c}_b(t)\mbs{\omega}^{ba}_b(t) + m_{s}\mbf{r}^{{sc}^\times}_b(t) \dot{\mbf{r}}^{sc}_b(t) ,\\
    \mbf{h}_b^{\mathcal{G}c/a}(t) &= \mbf{J}^{\mathcal{G}c}_b(t)\mbs{\omega}^{ba}_b(t) + m_{g}\mbf{r}^{{pc}^\times}_b(t) \dot{\mbf{r}}^{pc}_b(t).
\end{align}
The angular momentum of each RW $\mathcal{W}_i$ relative to point $c$ with respect to inertial reference frame $\mathcal{F}_{a}$ and resolved in $\mathcal{F}_b$ is given by
\beq
\mbf{h}_b^{\mathcal{W}_ic/a}(t) = \mbf{J}^{\mathcal{W}_ip}_b \dot{\mbs{\gamma}}^{\mathcal{W}_i}(t) + m_w \mbf{r}^{pc^\times}_b(t) \dot{\mbf{r}}^{pc}_b(t) + \mbf{J}^{\mathcal{W}_ic}_b(t) \dot{\mbs{\gamma}}^{\mathcal{W}_i}(t), \hspace{1em} i = 1, 2, 3,
\eeq
where $\mbf{J}_b^{\mathcal{W}_ic}(t) = \mbf{J}^{\mathcal{W}_ip}_b - m_w\mbf{r}_b^{{pc}^\times}(t)\mbf{r}_b^{{pc}^\times}(t)$ is the moment of inertia of each RW about point $c$, $\mbf{J}^{\mathcal{W}_i p}_b$ is the moment of inertia of each RW about point $p$, $m_w$ is the mass of a RW, and $\dot{\mbs{\gamma}}^{\mathcal{W}_i}(t)$ is the angular velocity of wheel $\mathcal{W}_i$ relative to frame $\mathcal{F}_b$, which is aligned with $\vect{b}^i$.
Specifically, $\dot{\mbs{\gamma}}^{\mathcal{W}_1}(t) = \Big[ \dot{\gamma}_1(t) \,\,\, 0 \,\,\, 0 \Big]^\trans$, $\dot{\mbs{\gamma}}^{\mathcal{W}_2}(t) = \Big[ 0 \,\,\, \dot{\gamma}_2(t) \,\,\, 0 \Big]^\trans$, $\dot{\mbs{\gamma}}^{\mathcal{W}_3}(t) = \Big[ 0 \,\,\, 0 \,\,\, \dot{\gamma}_3(t) \Big]^\trans$. 
Note that 
the RWs are assumed to be cylindrical, spinning about its axis of symmetry, and thus $ \mbf{J}^{\mathcal{W}_1 p}_b = \text{diag}(\f{m_w R^2}{2}, \f{m_w R^2}{4}, \f{m_w R^2}{4})$, $ \mbf{J}^{\mathcal{W}_2 p}_b = \text{diag}(\f{m_w R^2}{4}, \f{m_w R^2}{2}, \f{m_w R^2}{4})$, $ \mbf{J}^{\mathcal{W}_3 p}_b = \text{diag}(\f{m_w R^2}{4}, \f{m_w R^2}{4}, \f{m_w R^2}{2})$.
The time argument $(t)$ is omitted in the following for brevity.
Summing the angular momentum of three RWs relative to point $c$ results in
    \beq
        \sum^{3}_{i=1}\mbf{h}_b^{\mathcal{W}_i c/a} = \mbf{h}_b^{\text{RWs}} + 3 m_w \mbf{r}^{pc^\times}_b \dot{\mbf{r}}^{pc}_b + \sum^{3}_{i=1} \mbf{J}^{\mathcal{W}_ic}_b \mbs{\omega}^{ba}_b,
    \eeq
where $\mbf{h}_b^{\text{RWs}} = \sum_{i=1}^3 \mbf{J}^{\mathcal{W}_ip}_b \dot{\mbs{\gamma}}^{\mathcal{W}_i} = \f{m_w R^2}{2} \Big[ \dot{\gamma}_1 \,\,\, \dot{\gamma}_2 \,\,\, \dot{\gamma}_3 \Big]^\trans$ is the combination of three RWs' angular momentum relative to point $p$ with respect to inertial reference frame $\mathcal{F}_{a}$ resolved in frame $\mathcal{F}_{b}$, which is practically measurable based on the RW radius $R$ and spin rates $\dot{\gamma}_i$, $i = 1,2,3$.
The angular momentum of the whole spacecraft $\mathcal{B}$ relative to point $c$ and resolved in $\mathcal{F}_b$ is derived as
    \beq
    \label{eq:AngMomentumBody}
        \mbf{h}_b^{\mathcal{B}c/a} = \mbf{h}_b^{\mathcal{G}c/a} + \mbf{h}_b^{\mathcal{S}c/a} + \sum^{3}_{i=1} \mbf{h}_b^{\mathcal{W}_i c/a} 
        = \mbf{J}_b^{\mathcal{B}c}\mbs{\omega}^{ba}_b + \f{m_p^3 + m_s^3}{(m_p+m_s)^2}\mbf{r}^{{ps}^\times}_b \dot{\mbf{r}}^{ps}_b + \mbf{h}_b^{\text{RWs}}   .
    \eeq

The total torque applied to the spacecraft relative to point $c$ and resolved in $\mathcal{F}_b$ is $\mbs{\tau}_b^{\mathcal{B}c} = \mbs{\tau}_b^{\text{AMT}} + \mbs{\tau}_b^{\text{RCD}} + \mbs{\tau}_b^{\text{dist}}$, 
where $\mbs{\tau}_b^{\text{AMT}}$ is the torque caused by the SRP force due to an offset in CM and CP created by the AMT, $\mbs{\tau}_b^{\text{RCD}} = \Big[0,0,\tau_{b3}^{\text{RCD}}\Big]^\trans$ is the torque generated by the RCDs, and $\mbs{\tau}_b^{\text{dist}}$ is the disturbance torque.
Considering the SRP force $\mbf{f}_b^{\text{SRP}}$ acting on the geometric center of the sail membrane (point $s$) and resolved in $\mathcal{F}_b$, the resultant torque caused by the CM/CP offset is
    \beq
        \mbs{\tau}_b^{\text{AMT}} = \mbf{r}_b^{{sc}^\times} \mbf{f}_b^{\text{SRP}} = \f{m_s}{m_p+m_s}\mbf{r}_b^{ps^\times} \mbf{f}_b^{\text{SRP}}. \label{tau_AMT}
    \eeq

The equations of motion of the solar sail are derived beginning with Euler's Law for Rotation, given by
    \beq
        (\vect{{h}}^{\mathcal{B}c/a})^\fdot{a} = \vect{{\tau}}^{\mathcal{B}c} ,
    \eeq
    where $(\vect{{h}}^{\mathcal{B}c/a})^\fdot{a}$ is the time derivative of the solar sail's angular momentum with respect to the inertial frame $\mathcal{F}_a$. Applying the transport theorem to the expression results in
    \beq
    \label{eq:TransportTheoremEOM}
        (\vect{{h}}^{\mathcal{B}c/a})^\fdot{b} + \vect{{\omega}}^{ba} \times \vect{{h}}^{\mathcal{B}c/a}  = \vect{{\tau}}^{\mathcal{B}c} ,
    \eeq
where $(\vect{{h}}^{\mathcal{B}c/a})^\fdot{b}$ is the time derivative of the solar sail's angular momentum with respect to the body-fixed frame $\mathcal{F}_b$.
Substituting the expression for $\mbf{h}_b^{\mathcal{B}c/a}$ in~\eqref{eq:AngMomentumBody} into~\eqref{eq:TransportTheoremEOM}, the equations of motion representing the spacecraft attitude dynamics are derived as
\begin{multline}
    \mbf{J}_b^{\mathcal{B}c} \dot{\mbs{\omega}}_b^{ba} + \dot{\mbf{J}}_b^{\mathcal{B}c} \mbs{\omega}_b^{ba} + \mbs{\omega}_b^{ba^\times} \mbf{J}_b^{\mathcal{B}c} \mbs{\omega}_b^{ba} 
    + \f{m_p^3 + m_s^3}{(m_p+m_s)^2}  \mbs{\omega}_b^{ba^\times}\mbf{r}_b^{ps^\times}\dot{\mbf{r}}_b^{ps} \\+ \f{m_p^3 + m_s^3}{(m_p+m_s)^2}\mbf{r}_b^{ps^\times}\ddot{\mbf{r}}_b^{ps} + \mbs{\omega}_b^{ba^\times}\mbf{h}_b^{\text{RWs}} + \dot{\mbf{h}}_b^{\text{RWs}} = \mbs{\tau}_b^{\mathcal{B}c} . \label{EOM_1}
\end{multline}
Substituting in $\dot{\mbf{J}}_b^{\mathcal{B}} = - \f{m_p^3+m_s^3} {(m_p+m_s)^2} (\dot{\mbf{r}}_b^{ps^\times} \mbf{r}_b^{ps^\times} + \mbf{r}_b^{ps^\times} \dot{\mbf{r}}_b^{ps^\times} )$ and using the identities $\mbf{u}^\times \mbf{v} = -\mbf{v}^\times \mbf{u}$ and $\left(\mbf{u}^\times\mbf{v}\right)^\times = \mbf{u}^\times \mbf{v}^\times - \mbf{v}^\times \mbf{u}^\times$~\cite{DeRuiter2013},~\eqref{EOM_1} is rewritten as
\begin{multline}
    \mbf{J}_b^{\mathcal{B}c} \dot{\mbs{\omega}}_b^{ba} + \mbs{\omega}_b^{ba^\times} \mbf{J}_b^{\mathcal{B}c} \mbs{\omega}_b^{ba} + \mbs{\omega}_b^{ba^\times}\mbf{h}_b^{\text{RWs}} + \dot{\mbf{h}}_b^{\text{RWs}} \\
    + \f{m_p^3+m_s^3} {(m_p+m_s)^2} \bigg( \mbf{r}_b^{ps^\times}\ddot{\mbf{r}}_b^{ps} -2\dot{\mbf{r}}_b^{ps^\times} \mbf{r}_b^{ps^\times} \mbs{\omega}_b^{ba} \bigg) = \mbs{\tau}_b^{\mathcal{B}c} , \label{EOM_2}
\end{multline}
where $\mbs{\tau}_b^{\mathcal{B}c} = \mbs{\tau}_b^{\text{AMT}} + \mbs{\tau}_b^{\text{RCD}} + \mbs{\tau}_b^{\text{dist}} = \f{m_s}{m_p+m_s}\mbf{r}_b^{ps^\times} \mbf{f}_b^{\text{SRP}} + \mbs{\tau}_b^{\text{RCD}} + \mbs{\tau}_b^{\text{dist}}$, $\mbf{J}_b^{\mathcal{B}c}(t) = \mbf{J}^{\mathcal{P}p}_b + \mbf{J}^{\mathcal{S}s}_b - \f{m_p^3+m_s^3}{(m_p+m_s)^2}\mbf{r}_b^{{ps}^\times}(t)\mbf{r}_b^{{ps}^\times}(t)$, and $\mbf{r}^{ps}_b(t) = \Big[ r^{\text{AMT}}_{b1}(t) \,\,\, r^{\text{AMT}}_{b2}(t) \,\,\, {r}^{ps}_{b3} \Big]^\trans$. Note that the time-dependencies of $\mbf{J}_b^{\mathcal{B}c}(t)$ and $\mbf{r}^{ps}_b(t)$ are highlighted for the CM translation due to AMT actuation, while the time-dependencies of the remaining terms are omitted for brevity.


\subsection{RWs Attitude Control Law} \label{sec:RW_PID}
There exist many advanced control methods that can be implemented as RW controllers to meet the solar sail's attitude control requirements.
However, in this paper, a simple PID control law is chosen as the RW controller to perform fundamental attitude tracking based on the controller developed for Solar Cruiser~\cite{Inness2023}.
The time-derivative of the RWs angular momentum is directly associated to the torque acting on the spacecraft, where $\dot{\mbf{h}}^\text{RWs}_{b} = -\mbs{\tau}^{\text{RWs}}_b $. 
The RWs control torque synthesized with a PID control law is defined as
    \beq
        \mbs{\tau}^{\text{RWs}}_b = -\dot{\mbf{h}}^\text{RWs}_{b} = -\mbf{K}_p \mbstilde{\theta}(t) -\mbf{K}_d \dot{\mbstilde{\theta}}(t) -\mbf{K}_i \int^t_{t_0} \mbstilde{\theta}(\tau) \dee\tau , \label{PID_law}
    \eeq
where $\mbstilde{\theta}(t) = \mbs{\theta}(t) - \mbs{\theta}_d$, $\dot{\mbstilde{\theta}}(t) = \dot{\mbs{\theta}}(t) - \dot{\mbs{\theta}}_d $, and $\mbs{\theta}_d$, $ \dot{\mbs{\theta}}_d$ are the desired Euler angles and Euler angle rates of the desired trajectory, respectively. For the remainder of this paper, the desired attitude and angular rate are chosen to be $\mbf{0}$ for simplicity, although this is not a fundamental requirement of the proposed momentum management approach.

\section{MPC-Based Momentum Management Controller}\label{sec:MPC}

MPC is proposed to be a suitable strategy for solar sail momentum management system because of its ability to handle actuator constraints and optimize performance objectives simultaneously. The relatively long time scales involved in solar sail operation allow for large momentum management time steps, enabling less frequent actuation and computation. This reduced actuation frequency makes it potentially feasible to implement MPC onboard using standard, off-the-shelf QP solvers, provided that a linear prediction is used. 
The discrete-time linear dynamics are only used as the prediction model within MPC, while the inputs from the optimization solution are applied to the nonlinear system.

For demonstration purposes in this paper, knowledge of the spacecraft state, SRP force and the disturbance estimations are assumed to be perfect. Although this assumption is somewhat idealized, it is fair to assume that the solar sail's onboard flight computer provides an accurate state estimate and that the disturbance torque can be estimated using a Kalman filter, similar to the density estimation approach in~\cite{hayes2022atmospheric}. 

Section~\ref{sec:Linearization} presents the derivation of the system’s linearized dynamics about the operational state of interest, which is not necessarily an equilibrium point, to facilitate the linear time-invariant (LTI) prediction model for the MPC formulation.
The AMT and RCDs exhibit distinct actuation mechanisms that are explicitly accounted for in this framework. The AMT operates continuously, translating the spacecraft bus with a bounded speed, while the RCDs generate discrete on-off torque pulses. Given the large momentum management time step relative to the RWs control time step, improper discretization could lead to significant inaccuracies and misrepresentation of actuator inputs. To better model these momentum management inputs, tailored discretization schemes are employed, with FOH applied to the AMT’s continuous motion and ZOH used for the RCDs' on-off inputs. 
The discretization and quantization methodologies are detailed in Sections~\ref{sec:Discretization} and~\ref{sec:PWM}. 
To ensure the RWs maintain their primary role in spacecraft attitude tracking while the momentum management system focuses on mitigating RW angular momentum buildup, a slack variable is introduced in the MPC cost function. This variable penalizes RW angular momentum growth, directing AMT movement and RCD pulses toward momentum management tasks. Additional penalties are applied to minimize AMT motion, as maintaining a fixed AMT position is less costly than a large translation of mass. The complete MPC policy formulation is presented in Section~\ref{sec:MPC_formulation}. 
While PWM-based quantization is applied to the RCDs’ inputs after the MPC optimization is solved, this introduces a mismatch between the optimized input and the actual applied input to the dynamic propagation. To address this, an iterative backwards-in-time MPC approach is proposed in Section~\ref{sec:iter_back}, which incorporates the effects of quantized RCD torque into the optimization process. 


\subsection{Linear Prediction Model} \label{sec:Linearization}

Turning the MPC into a QP is a powerful tool that enables the capability to implement the algorithm on board and in real-time.
To achieve this, a discrete-time linear dynamic model is needed. 
Consider the perturbed state and input around the operation point of interest, $\mbs{\theta} = \bar{\mbs{\theta}} + \delta\mbs{\theta}$, $\mbs{\omega}^{ba}_b = \bar{\mbs{\omega}}^{ba}_b + \delta\mbs{\omega}^{ba}_b$, $\mbf{h}^{\text{RWs}}_b = \mbfbar{h}^{\text{RWs}}_b + \delta\mbf{h}^{\text{RWs}}_b$, $\mbf{r}^{ps}_b = \mbfbar{r}^{ps}_b + \delta \mbf{r}^{ps}_b$, $\mbs{\tau}^{\text{RCD}}_{b} = \bar{\mbs{\tau}}^{\text{RCD}}_{b} + \delta \mbs{\tau}^{\text{RCD}}_{b}$, and $\mbf{e}^\text{int} = \mbfbar{e}^\text{int} + \delta \mbf{e}^\text{int}$, where
$ \mbf{e}^\text{int} = \int^t_{t_0} (\mbs{\theta}(\tau) - \mbs{\theta}_d) \dee\tau$ is the internal state representing the integral term of PID law.

The bar notation on the state and input represents the operation point to be linearized about, and the $\delta$ notation represents perturbations from this operating point.
The perturbed disturbance is given by $\mbs{\tau}^{\text{dist}}_b = \bar{\mbs{\tau}}^{\text{dist}}_b +\delta \mbs{\tau}^{\text{dist}}_b$.
Given the perturbed state and input, 
their time derivatives are given by $ \dot{\mbs{\theta}} = \dot{\bar{\mbs{\theta}}} + \delta\dot{\mbs{\theta}} $, $ \dot{\mbs{\omega}}_b^{ba} = \dot{\bar{\mbs{\omega}}}_b^{ba} + \delta\dot{\mbs{\omega}}_b^{ba} $, $\dot{\mbf{h}}^{\text{RWs}}_b = \dot{\mbfbar{h}}^{\text{RWs}}_b + \delta \dot{\mbf{h}}^{\text{RWs}}_b$, $\dot{\mbf{e}}^\text{int} = \dot{\mbfbar{e}}^\text{int} + \delta \dot{\mbf{e}}^\text{int}$, $\dot{\mbf{r}}^{ps}_b = \dot{\mbfbar{r}}^{ps}_b + \delta \dot{\mbf{r}}^{ps}_b$.

The spacecraft's angular velocity $\mbs{\omega}_b^{ba}= \mbf{S}(\mbs{\theta})\dot{\mbs{\theta}}$ as derived in Section~\ref{Sec:nonlin_dyn} under this  small perturbation becomes
    \beq
        \delta\mbs{\omega}^{ba}_b = \bbm \delta\omega_1 \\ \delta\omega_2 \\ \delta\omega_3 \ebm = \bbm 1 & 0 & -\delta\theta_2\\ 0 & 1 & \delta\theta_1  \\ 0 & -\delta\theta_1 & 1 \ebm \bbm \delta\dot{\theta}_1 \\ \delta\dot{\theta}_2 \\ \delta\dot{\theta}_3 \ebm .
    \eeq
Dropping higher-order terms in this expression leads to the approximation 
\beq
    \delta\mbs{\omega}^{ba}_b \approx \delta \dot{\mbs{\theta}} . \label{eq:linEOM_p1}
\eeq
The desired trajectory is chosen to be zero, such that $\mbs{\theta}_d = \mbf{0}$, $\dot{\mbs{\theta}}_d = \mbf{0}$, and the RW PID torque becomes
\begin{equation}
    \delta\dot{\mbf{h}}_b^{\text{RWs}} = -\delta\mbs{\tau}^{\text{RWs}}_b 
    = \mbf{K}_p \delta\mbs{\theta} +\mbf{K}_d \delta\dot{\mbs{\theta}} +\mbf{K}_i \delta \mbf{e}^\text{int} . \label{eq:linEOM_p2}
\end{equation}
The time-derivative of the integral term is given by
\beq
    \delta \dot{\mbf{e}}^\text{int} = \delta \mbs{\theta} . \label{eq:linEOM_p3}
\eeq
While the future input values of $\ddot{\mbf{r}}^{ps}_b$, $\dot{\mbf{r}}^{ps}_b$ and $\mbs{\tau}^{\text{RCD}}_{b}$ are unknown, the  nominal inputs $\ddot{\mbfbar{r}}^{ps}_b = \mbf{0}$, $\dot{\mbfbar{r}}^{ps}_b = \mbf{0}$ and $\bar{\mbs{\tau}}^{\text{RCD}}_{b} = \mbf{0}$ are used in the prediction model. The perturbed input rate (AMT velocity) is also assumed zero, $\delta \ddot{\mbf{r}}^{ps}_b = \mbf{0}$, $\delta \dot{\mbf{r}}^{ps}_b = \mbf{0}$.
Substituting the perturbed states, their derivatives, and inputs into~\eqref{EOM_2} and removing higher-order terms results in
\begin{multline}
\label{eq:EOM_TaylorSeries}
        \mbfbar{J}_b^{\mathcal{B}c} \big(\dot{\mbsbar{\omega}}_b^{ba} +\delta\dot{\mbs{\omega}}_b^{ba}\big) +\f{m_p^3+m_s^3}{(m_p+m_s)^2} \Big(\mbfbar{r}^{ps^\times}_b\dot{\mbsbar{\omega}}_b^{ba^\times} + \big( \mbfbar{r}^{ps^\times}_b\dot{\mbsbar{\omega}}_b^{ba}\big)^\times \Big) \delta\mbf{r}^{ps}_b
       \\ = -\mbsbar{\omega}_b^{ba^\times} \big(\mbfbar{J}_b^{\mathcal{B}c}\mbsbar{\omega}_b^{ba} +\mbfbar{h}^{\text{RWs}}_b \big)
        +\f{m_s}{m_p+m_s}\mbfbar{r}^{ps^\times}_b\mbf{f}_b^{\text{SRP}} +\mbs{\tau}^{\text{dist}}_b +\delta \mbs{\tau}^{\text{dist}}_b -\dot{\mbfbar{h}}_b^{\text{RWs}} -\delta\dot{\mbf{h}}_b^{\text{RWs}}  
        \\ -\f{m_p^3+m_s^3}{(m_p+m_s)^2}\mbsbar{\omega}_b^{ba^\times} \big( \mbfbar{r}^{ps^\times}_b \delta\mbf{r}^{ps^\times}_b +\delta\mbf{r}^{ps^\times}_b \mbfbar{r}^{ps^\times}_b\big) \mbsbar{\omega}_b^{ba}  +\f{m_s}{m_p+m_s}\delta\mbf{r}^{ps^\times}_b\mbf{f}_b^{\text{SRP}} 
        \\ -\delta\mbs{\omega}_b^{ba^\times} \mbfbar{J}_b^{\mathcal{B}c} \mbsbar{\omega}_b^{ba}  -\mbsbar{\omega}_b^{ba^\times} \mbfbar{J}_b^{\mathcal{B}c} \delta\mbs{\omega}_b^{ba} -\delta\mbs{\omega}_b^{ba^\times}\mbfbar{h}_b^{\text{RWs}} -\mbsbar{\omega}_b^{ba^\times}\delta\mbf{h}_b^{\text{RWs}} +\delta\mbs{\tau}^{\text{RCD}}_b ,
\end{multline}
where $\mbfbar{J}_b^{\mathcal{B}c} = \mbf{J}_b^{\mathcal{P}p}+\mbf{J}_b^{\mathcal{S}s} -\f{m_p^3+m_s^3}{(m_p+m_s)^2} \mbfbar{r}^{ps^\times}_b\mbfbar{r}^{ps^\times}_b $.
Considering $\ddot{\mbfbar{r}}^{ps}_b =\mbf{0}$, $\dot{\mbfbar{r}}^{ps}_b = \mbf{0}$, $\mbsbar{\tau}^{\text{RCD}}_{b} = \mbf{0}$, and substituting the nominal operating points into the full dynamics in equation~\eqref{EOM_2}, results in the nominal dynamics
\beq
\label{eq:NominalEOM}
    \mbfbar{J}_b^{\mathcal{B}c}\dot{\mbsbar{\omega}}_b^{ba} = -\mbsbar{\omega}_b^{ba^\times}\mbfbar{J}_b^{\mathcal{B}c}\mbsbar{\omega}_b^{ba} -\mbsbar{\omega}_b^{ba^\times}\mbfbar{h}^{\text{RWs}}_b +\f{m_s}{m_p+m_s}\mbfbar{r}^{ps^\times}_b\mbf{f}_b^{\text{SRP}} +\mbsbar{\tau}^{\text{dist}}_b  -\dot{\mbfbar{h}}_b^{\text{RWs}}.
\eeq
The nominal dynamics of~\eqref{eq:NominalEOM} are used to cancel out equivalent terms on both sides of~\eqref{eq:EOM_TaylorSeries}, resulting in the linearized equations of motion
\begin{multline}
        \mbfbar{J}_b^{\mathcal{B}c} \delta\dot{\mbs{\omega}}_b^{ba} = 
        -\f{m_p^3+m_s^3}{(m_p+m_s)^2}\mbsbar{\omega}_b^{ba^\times} \Big(\mbfbar{r}^{ps^\times}_b\mbsbar{\omega}_b^{ba^\times} + \big( \mbfbar{r}^{ps^\times}_b\mbsbar{\omega}_b^{ba}\big)^\times \Big) \delta\mbf{r}^{ps}_b 
        \\ -\f{m_p^3+m_s^3}{(m_p+m_s)^2} \Big(\mbfbar{r}^{ps^\times}_b\dot{\mbsbar{\omega}}_b^{ba^\times} + \big( \mbfbar{r}^{ps^\times}_b\dot{\mbsbar{\omega}}_b^{ba}\big)^\times \Big) \delta\mbf{r}^{ps}_b \\
        -\f{m_s}{m_p+m_s}\mbf{f}_b^{\text{SRP}^\times}\delta\mbf{r}^{ps}_b 
        +\Big( - \mbsbar{\omega}_b^{ba^\times}\mbfbar{J}_b^{\mathcal{B}c} +\big( \mbfbar{J}_b^{\mathcal{B}c} \mbsbar{\omega}_b^{ba} \big)^\times +\mbfbar{h}_b^{\text{RWs}^\times} \Big) \delta\mbs{\omega}_b^{ba} -\mbsbar{\omega}_b^{ba^\times}\delta\mbf{h}_b^{\text{RWs}} \\
        -\f{m_p^3+m_s^3}{(m_p+m_s)^2} \Big(\mbfbar{r}^{ps^\times}_b\dot{\mbsbar{\omega}}_b^{ba^\times} + \big( \mbfbar{r}^{ps^\times}_b\dot{\mbsbar{\omega}}_b^{ba}\big)^\times \Big) \delta\mbf{r}^{ps}_b
        +\delta\mbs{\tau}^{\text{RCD}}_b +\delta \mbs{\tau}^{\text{dist}}_b -\delta\dot{\mbf{h}}_b^{\text{RWs}}, \label{eq:linearized_EOMs}
\end{multline}
where $\dot{\mbsbar{\omega}}_b^{ba} = \mbfbar{J}_b^{\mathcal{B}c^{-1}}\big( -\mbsbar{\omega}_b^{ba^\times}\mbfbar{J}_b^{\mathcal{B}c}\mbsbar{\omega}_b^{ba} -\mbsbar{\omega}_b^{ba^\times}\mbfbar{h}^{\text{RWs}}_b +\f{m_s}{m_p+m_s}\mbfbar{r}^{ps^\times}_b\mbf{f}_b^{\text{SRP}} +\mbsbar{\tau}^{\text{dist}}_b  -\dot{\mbfbar{h}}_b^{\text{RWs}} \big)$ from the nominal dynamics in~\eqref{eq:NominalEOM}.
It is assumed that a full-state measurement $\mbf{y}(t) = \mbf{x}(t)$ is accessible from the sensors on board without any measurement noise.
With equations~\eqref{eq:linEOM_p1},~\eqref{eq:linEOM_p2},~\eqref{eq:linEOM_p3}, and~\eqref{eq:linearized_EOMs} a linear state-space realization of the solar sail's dynamics are given by
\begin{align}
    \dot{\mbf{x}}(t) &= \mbf{A}(t)\mbf{x}(t) + \mbf{B}_w(t)\mbf{w}(t) + \mbf{B}_{u1}(t)\mbf{u}^\text{AMT}(t) +  \mbf{B}_{u2}(t)u^\text{RCD}(t), \label{eq:lin_EOM_ss} \\
    \mbf{y}(t) &= \mbf{x}(t), \label{eq:lin_EOM_output}
\end{align}
where $\mbf{x}(t) = \Big[ \delta\mbs{\theta}^\trans \,\,\, \delta{\mbs{\omega}^{ba}_b}^\trans \,\,\, \delta\mbf{h}^{\text{RWs}^\trans}_b \,\,\,  \delta\mbf{e}^{\text{int}^\trans} \Big]^\trans$, 
$\mbf{w}(t) = \delta \mbs{\tau}^{\text{dist}}_b$, $u^\text{RCD}(t) = \delta \tau^{\text{RCD}}_{b3}$, $\mbf{u}^\text{AMT}(t) = \Big[ \delta r^\text{AMT}_{b1} \,\,\, \delta r^\text{AMT}_{b2} \Big]^\trans$,
\bdis
    \mbf{A}(t) = \bbm \mbf{0}_{3 \times 3} &\mbf{1}_{3 \times 3} &\mbf{0}_{3 \times 3} & \mbf{0}_{3 \times 3} \\ 
        -\mbfbar{J}_b^{\mathcal{B}c^{-1}}\mbf{K}_{p} &\f{\p \mbf{f}_2}{\p\mbs{\omega}_b^{ba}}\Bigg|_{\bar{\mbf{x}}, \bar{\mbf{u}}} & -\mbfbar{J}_b^{\mathcal{B}c^{-1}}\mbsbar{\omega}_b^{ba^\times} & -\mbfbar{J}_b^{\mathcal{B}c^{-1}}\mbf{K}_{i} \\ 
        \mbf{K}_{p} &\mbf{K}_{d} &\mbf{0}_{3 \times 3}&\mbf{K}_{i}  \\ 
        \mbf{1}_{3 \times 3}&\mbf{0}_{3 \times 3}&\mbf{0}_{3 \times 3}&\mbf{0}_{3 \times 3} \ebm,
\edis
\bdis
        \mbf{B}_w(t) = \bbm \mbf{0}_{3 \times 3}\\ \mbfbar{J}_b^{\mathcal{B}c^{-1}} \\ \mbf{0}_{3 \times 3} \\\mbf{0}_{3 \times 3}  \ebm ,
    \hspace{1em} \mbf{B}_{u1}(t) = \bbm \mbf{0}_{3 \times 2} \\ \f{\p \mbf{f}_2}{\p\mbf{r}_b^{ps}}\Bigg|_{\bar{\mbf{x}}, \bar{\mbf{u}}}\bbm \mbf{1}_{2 \times 2}\\ \mbf{0}_{1 \times 2} \ebm  \\ \mbf{0}_{3 \times 2} \\ \mbf{0}_{3 \times 2} \ebm,
    \hspace{1em} \mbf{B}_{u2}(t) = \bbm \mbf{0}_{3 \times 3} \\ \mbfbar{J}_b^{\mathcal{B}c^{-1}} \\ \mbf{0}_{3 \times 3} \\ \mbf{0}_{3 \times 3} \ebm \bbm 0\\0\\1\ebm,
\edis
\begin{multline*}
\f{\p \mbf{f}_2}{\p\mbf{r}_b^{ps}}\Big|_{\bar{\mbf{x}}, \bar{\mbf{u}}} = \mbfbar{J}_b^{\mathcal{B}c^{-1}} \Big( -\f{m_p^3+m_s^3}{(m_p+m_s)^2}\mbsbar{\omega}_b^{ba^\times} \Big(\mbfbar{r}^{ps^\times}_b\mbsbar{\omega}_b^{ba^\times} + \big( \mbfbar{r}^{ps^\times}_b \mbsbar{\omega}_b^{ba} \big)^\times \Big)  \\-\f{m_p^3+m_s^3}{(m_p+m_s)^2} \Big(\mbfbar{r}^{ps^\times}_b\dot{\mbsbar{\omega}}_b^{ba^\times} + \big( \mbfbar{r}^{ps^\times}_b\dot{\mbsbar{\omega}}_b^{ba}\big)^\times \Big) -\f{m_s}{m_p+m_s}\mbf{f}_b^{\text{SRP}^\times} \Big),
\end{multline*}
\bdis
\f{\p \mbf{f}_2}{\p\mbs{\omega}_b^{ba}}\Big|_{\bar{\mbf{x}}, \bar{\mbf{u}}}  = \mbfbar{J}_b^{\mathcal{B}c^{-1}} \Big( \big(\mbfbar{J}_b^{\mathcal{B}c}\mbsbar{\omega}_b^{ba}\big)^\times -\mbsbar{\omega}_b^{ba^\times}\mbfbar{J}_b^{\mathcal{B}c} +\mbfbar{h}^{\text{RWs}^\times}_b -\mbf{K}_d
        \Big),
\edis 
$\bar{\mbf{x}} = \Big[ \mbsbar{\theta}^\trans \,\,\, {\mbsbar{\omega}^{ba^\trans}_b} \,\,\, \mbfbar{h}^{\text{RWs}^\trans}_b \,\,\,  \mbfbar{e}^{\text{int}^\trans} \Big]^\trans$, $\bar{\mbf{u}} = \Big[ \mbsbar{\tau}^{\text{dist}^\trans}_b \,\,\, \mbfbar{u}^{\text{AMT}^\trans} \,\,\,  \bar{\tau}^{\text{RCD}}_{b3} \Big]^\trans$, and the time-dependencies in the state-space matrices are omitted for brevity. 
It is worth noting that this is a linear time-varying (LTV) state-space model, where the matrices $\mbf{A}(t)$ and $\mbf{B}(t) = \Big[ \mbf{B}_w(t) \,\,\, \mbf{B}_{u1}(t) \,\,\, \mbf{B}_{u2}(t) \Big]$ include time-dependent parameters $\mbsbar{\omega}_b^{ba^\times}(t)$, $\mbfbar{h}^{\text{RWs}}_b(t)$, $\mbfbar{r}^{ps}_b(t)$, and $\mbsbar{\tau}^{\text{dist}}_b(t)$, which are part of the state $\mbfbar{x}(t)$ and input $\mbfbar{u}(t)$ to be linearized about.
The term $\mbfbar{J}_b^{\mathcal{B}c}(t)$ is also time-dependent due to the CM translation associated with $\mbfbar{r}^{ps}_b(t)$.

For the purpose of implementing an linear-quadratic (LQ)-MPC policy efficiently as a QP, the LTV dynamics are simplified as an LTI prediction model within the proposed MPC algorithm, where the time-dependent parameters are only updated with the latest measurement at every momentum management time step, and these parameters are kept constant throughout the MPC prediction horizon.
This approximation improves the computation efficiency of the controller and, if successful, demonstrates the robustness of the controller even when the prediction model is not exactly accurate.
Specifically, it is considered that $\mbf{A}(t) = \mbf{A}(t_k)$ and $\mbf{B}(t) = \mbf{B}(t_k)$, where the time varying parameters $\mbsbar{\omega}^{ba}_b(t_k)$, ${\mbfbar{h}^{\text{RWs}}_{b}}(t_k)$, $\mbfbar{r}^{ps}_b(t_k)$, $\mbsbar{\tau}^{\text{dist}}_b(t_k)$ are updated at momentum management time step $t = t_k$, and stay constant throughout the entire prediction horizon in the MPC optimization algorithm at time $t = t_k$.
The SRP force $\mbf{f}^{\text{SRP}}_b$ is assumed constant,  as it is assumed that the spacecraft attitude is changing slowly or held at some fixed orientation relatively to the Sun for the duration of the prediction horizon, allowing for the matrix $\mbf{B}_{u1}(t_k)$ to be held as a constant.


\subsection{Discretization - Mixed FOH and ZOH Tailored for Momentum Management Actuators} \label{sec:Discretization}

A discrete-time model is used as the MPC policy to forsee the propagation of the system's state over a finite prediction horizon.
There are different methods that can be used to discretize a continuous-time model into a discrete-time model depending on the required accuracy, complexity, and numerical efficiency.
Given the distinct actuation characteristics of the solar sail’s momentum management actuators, this section outlines a tailored discretization strategy that employs a FOH for the continuous AMT input and a ZOH for the on-off RCD input. This hybrid discretization approach captures the unique attributes of each actuator while preserving the fidelity required for the MPC formulation.

The AMT adjusts the solar sail's CM within a plane relative to the CP, inducing torques in the pitch and yaw axes. These torques can trim disturbances, manage angular momentum, or contribute to attitude control. AMT actuation constraints include the range and rate of translation, which necessitate a discretization method capable of capturing continuous motion over the longer momentum management time steps. 
A FOH discretization is employed to model the AMT’s continuous movement, providing linear interpolation of CM/CP shifts between successive momentum management time steps. This approach more accurately reflects the gradual translation of the AMT compared to a ZOH, which assumes instantaneous jumps between positions. In the MPC optimization problem, displacement limits and rate constraints on AMT inputs are incorporated, ensuring physically realizable solutions.
The RCDs operate as discrete pulse-based actuators, generating torques through reflectivity modulation. Since RCD inputs are inherently discrete, ZOH is used to approximate the constant torque output over each momentum management time step. This method ensures compatibility with the integer constraints introduced by the RCD actuation mechanism. 
The remainder of this section presents the proposed tailored hybrid discretization approach.

As a first step, consider a scalar example of a linear interpolation-based FOH. The continuous-time input $u(t)$ can be represented between discrete times $t_{k}$ and $t_{k+1}$ as
\beq \label{eq:FOH_u}
    u(t) =  \lambda^-_k(t) u_k + \lambda^+_k(t) u_{k+1},  \quad t \in [t_{k}, t_{k+1}],       
\eeq        
where $u_k = u(t_k)$, $u_{k+1} = u(t_{k+1})$, and the scaling parameters $ \lambda^-_k(t) = \f{t_{k+1}-t}{t_{k+1}-t_k} $ and $ \lambda^+_k(t) = \f{t-t_{k}}{t_{k+1}-t_k} $ linearly interpolate the input. These scaling parameters are used in the subsequent FOH discretization scheme. 
To proceed with the discretization, the linearized dynamics in~\eqref{eq:lin_EOM_ss}
are reorganized as
\beq
    \dot{\mbf{x}}(t)  = \mbf{A}(t)\mbf{x}(t) +\mbf{B}_w(t)\mbf{w}(t) +\mbf{B}_u(t)\mbf{u}(t) ,
\eeq
where $\mbf{B}_u(t) = \Big[ \mbf{B}_{u1}(t) \,\,\, \mbf{B}_{u2}(t) \Big]$ and $\mbf{u}(t) = \Big[ \mbf{u}^{\text{AMT}^\trans}(t) \,\,\, u^{\text{RCD}}(t)\Big]^\trans$.
At every discrete momentum management time step $t_k$, the continuous-time Jacobian matrices derived in Section~\ref{sec:Linearization} are evaluated with the current states, where $\mbf{A}_{\ell} = \mbf{A}(t_k)$, $\mbf{B}_{w,\ell} = \mbf{B}_w(t_k)$, and $\mbf{B}_{u,\ell} = \Big[ \mbf{B}_{u1}(t_k) \,\,\, \mbf{B}_{u2}(t_k) \Big] $. 
It is approximated that $\mbf{A}_{\ell}$, $\mbf{B}_{w,\ell}$, and $\mbf{B}_{u,\ell}$ remain constant throughout the prediction horizon, where the operation point $\mbsbar{\omega}^{ba}_b(t)$, ${\mbfbar{h}^{\text{RWs}}_{b}}(t)$, and $\mbfbar{r}^{ps}_b(t)$ are evaluated at time $t_k$. 
Considering a single momentum management time step $t_k \leq t \leq t_{k+1}$, the discrete-time state solution at $t_{k+1}$ is
\begin{multline} \label{eq:int_x_solution}
    \mbf{x}_{k+1} = e^{\mbf{A}_\ell(t_{k+1} -t_k)} \mbf{x}_k + \int^{t_{k+1}}_{t_k} e^{\mbf{A}_\ell(t_{k+1}-\tau)}\mbf{B}_{w,\ell}\mbf{w}(\tau)\dee\tau \\+ \int^{t_{k+1}}_{t_k} e^{\mbf{A}_\ell(t_{k+1}-\tau)}\mbf{B}_{u,\ell}\mbf{u}(\tau)\dee\tau ,
\end{multline}
where $\mbf{x}_k = \mbf{x}(t_k)$ and $\mbf{x}_{k+1} = \mbf{x}(t_{k+1})$.
A collective input $\mbf{u}(\tau)$ formulated by the linear interpolation in equation~\eqref{eq:FOH_u} is written as
\begin{align}
    \mbf{u}(\tau) &=  \mbs{\Lambda}^-_k(\tau) \mbf{u}_k + \mbs{\Lambda}^+_k(\tau) \mbf{u}_{k+1} \nonumber\\
    &= \bbm \f{t_{k+1}-\tau}{t_{k+1}-t_k}\mbf{1}&\mbf{0}\\\mbf{0}&1\ebm \bbm \mbf{u}^\text{AMT}_{k}\\u^\text{RCD}_{k} \ebm + \bbm \f{\tau-t_{k}}{t_{k+1}-t_k}\mbf{1}&\mbf{0}\\\mbf{0}&0\ebm \bbm \mbf{u}^\text{AMT}_{k+1}\\u^\text{RCD}_{k+1} \ebm, \label{eq:input_DTLTI} 
\end{align}
where the entries of $\mbs{\Lambda}^-_k(\tau)$ and $\mbs{\Lambda}^+_k(\tau)$ are chosen such that 
the input $\mbf{u}^\text{AMT}$ uses a FOH, and the input $u^\text{RCD}$ uses a ZOH. The disturbance $\mbf{w}(\tau) = \mbf{w}_k$ is discretized using a ZOH. 
Although this discretization implementing with the coexistence of ZOH and FOH inputs is not typical, it provides a solution that is well-tailored to the specific actuator properties of the AMT and RCDs.

With the proposed discretization scheme, the discrete-time state solution at $t_{k+1}$ is found by substituting~\eqref{eq:input_DTLTI} into~\eqref{eq:int_x_solution}, resulting in
\beq \label{eq:DT_dynamics_ss}
    \mbf{x}_{k+1} = \mbf{A}_{k}\mbf{x}_k + \mbf{B}_{w,k}\mbf{w}_k + \mbf{B}_{u,k}^{-}\mbf{u}_k + \mbf{B}_{u,k}^{+}\mbf{u}_{k+1} ,
\eeq
where $\mbf{u}_k = \Big[ \mbf{u}^{\text{AMT}^\trans}_{k} \,\,\, u^{\text{RCD}}_{k} \Big]^\trans$, 
$\mbf{A}_{k} = e^{\mbf{A}_\ell(t_{k+1} -t_k)}$, $\mbf{B}_{w,k} = \int^{t_{k+1}}_{t_k} e^{\mbf{A}_\ell(t_{k+1}-\tau)}\mbf{B}_{w,\ell}\dee\tau$, 
$
\mbf{B}_{u,k}^{-} = \int^{t_{k+1}}_{t_k} e^{\mbf{A}_\ell(t_{k+1}-\tau)}\mbf{B}_{u,\ell}\mbs{\Lambda}^-_k(\tau)\dee\tau$, and $\mbf{B}_{u,k}^{+} = \int^{t_{k+1}}_{t_k} e^{\mbf{A}_\ell(t_{k+1}-\tau)}\mbf{B}_{u,\ell}\mbs{\Lambda}^+_k(\tau)\dee\tau$. These LTV state-space matrices can be solved for as $\mbf{A}_{k} = \mbs{\Phi}_x(t_{k+1},t_k)$, $\mbf{B}_{w,k} = \mbs{\Phi}_w(t_{k+1},t_k)$, $\mbf{B}_{u,k}^{-} = \mbs{\Phi}_u^{-}(t_{k+1},t_k)$, $\mbf{B}_{u,k}^{+} = \mbs{\Phi}_u^{+}(t_{k+1},t_k)$, through the numerical integration of the matrix differential equations 
\begin{align}
\dot{\mbs{\Phi}}_x(t,t_k) &= \mbf{A}_{\ell} \mbs{\Phi}_x(t,t_k), \\
\dot{\mbs{\Phi}}_w(t,t_k) &= \mbf{A}_{\ell}\mbs{\Phi}_w(t,t_k) + \mbf{B}_{w,\ell}, \\ 
\dot{\mbs{\Phi}}_u^{-}(t,t_k) &= \mbf{A}_{\ell}\mbs{\Phi}_u^{-}(t,t_k) + \mbf{B}_{u,\ell} \mbs{\Lambda}^-_k, \\
\dot{\mbs{\Phi}}_u^{+}(t,t_k) &= \mbf{A}_{\ell}\mbs{\Phi}_u^{+}(t,t_k) + \mbf{B}_{u,\ell} \mbs{\Lambda}^+_k,
\end{align}
with initial values $\mbs{\Phi}_x(t_{k},t_k) = \mbf{1}_{n_x}$, $\mbs{\Phi}_w(t_{k},t_k) = \mbf{0}_{n_x \times n_w}$, $\mbs{\Phi}_u^{-}(t_{k},t_k) = \mbf{0}_{n_x \times n_u}$, and $\mbs{\Phi}_u^{+}(t_{k},t_k) = \mbf{0}_{n_x \times n_u}$ over the time interval $t \in [t_{k}, t_{k+1}]$. 
The prediction model used by the proposed MPC algorithm is then chosen as the discrete-time LTV model in~\eqref{eq:DT_dynamics_ss} evaluated at time $t_k$. 
Since the MPC prediction model accuracy significantly affects the performance, this approach is specifically tailored to account for the significantly different actuation natures of AMT and RCDs, where a FOH discretization is used to capture the smooth continuous translation of AMT and a ZOH discretization is used to model the discrete on-off pulsing of RCDs.

\subsection{PWM-Based Quantization of Integer RCDs Torque and Actuation Thresholds} \label{sec:PWM}

The ZOH discretization of the RCD input, as described in Section~\ref{sec:Discretization}, allows for continuous input values within $-u^\text{RCD}_\text{max} \leq u^\text{RCD}_k \leq u^\text{RCD}_\text{max}$, where a negative $u^\text{RCD}_k$ represents a counterclockwise torque about the roll axis, and a positive $u^\text{RCD}_k$ provides a clockwise torque about the roll axis. 
In practice, the RCDs are activated through small electrical power inputs that adjust their reflectivity, producing discrete, fixed-magnitude torque pulses about the roll axis. These pulses result in on-off actuation, where at any instance in time, the RCD torque generated is represented by $\alpha_\text{on-off}u^\text{RCD}_\text{max}$ with $\alpha_\text{on-off} \in \{-1, 0, 1\}$ and $u^\text{RCD}_\text{max} = \tau^\text{RCD}_{b3,\text{on}}$. This quantized actuation behavior complicates real-time optimization for MPC, as it results in a mixed-integer programming (MIP) problem, which is non-convex and has a computational expense that is prohibitive for onboard applications.

To avoid the need to solve an MIP, heuristic approaches can be implemented that address the on-off actuation of the RCDs in a computationally-efficient manner. A simple option is to directly round continuous MPC input values to the nearest integer, with a fixed on pulse spanning the duration of the full momentum management time step. Although this approach is simple, it can potentially lead to a substantial difference between the predicted system response with MPC and the actual system response. Another option involves a PWM approach inspired by the electric thruster control quantization method in~\cite{zlotnik2017mpc}, where the continuous RCD input value is converted to a set of fixed-magnitude, on-off pulses whose durations are chosen in a manner to recreate the effect of the continuous RCD input value obtained from MPC. 
Although in general multiple PWM on-off pulses can be implemented in a single discrete timestep $\Delta t$, in this particular application it is desirable to reduce the number of on-off cycles as much as possible and only a single on-off cycle is chosen per timestep. To this end, we propose a PWM-inspired quantization scheme where the RCD turns on at the beginning of the discrete timestep and then has a pulse length of
\beq
    t_c = \Delta t \cdot \f{u^\text{RCD}_{k,\text{mpc}}}{u^\text{RCD}_\text{max}} ,
\eeq
where $u^\text{RCD}_{k,\text{mpc}}$ is the optimal continuous RCD input determined from MPC at the $k^{\text{th}}$ timestep, and $\Delta t = t_{k+1} - t_k$ is the length of timestep. The RCD pulse starts at time $t_k$ and cuts off at time $t_k+t_c$, where $t_c \leq \Delta t$.

\begin{figure}[h]
\centering
\subfigure[MPC optimal input.]
{
		\includegraphics[width=0.48\textwidth]{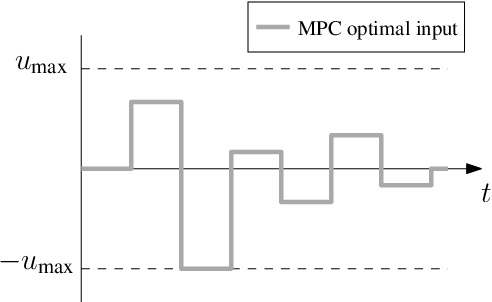} %
		\label{Fig_pwm_1}
}
\subfigure[PWM quantized input.]
{
		\includegraphics[width=0.48\textwidth]{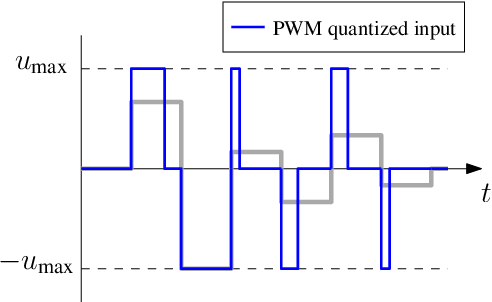} %
		\label{Fig_pwm_2}
}
\subfigure[PWM quantized input with threshold.]
{
		\includegraphics[width=0.48\textwidth]{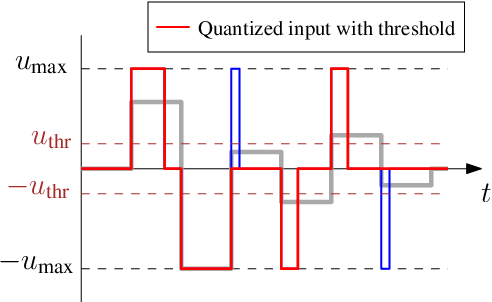} %
		\label{Fig_pwm_3}
}
\centering
\vspace{-8pt}
\caption{Quantization of the RCD input using a PWM-inspired approach and thresholding.}\label{Fig:PWM}
\end{figure}

Given that frequent actuation, such as a large number of on-off RCD cycles, 
shortens actuator lifespan, 
reducing the number of non-zero control inputs is desirable. In this work, the amount of actuator usage is considered a performance metric for evaluating momentum management strategies. Inspired by the Solar Cruiser momentum management approach, where the actuator's activation is governed by a threshold policy based on RW angular momentum bounds~\cite{Inness2023}, a similar thresholding strategy is employed here.
To avoid unnecessary short on-off RCD cycles, a threshold value $u_\text{thr}$ is chosen to define a deadband, where any continuous MPC input satisfying $|u^\text{RCD}_{k,\text{mpc}}| < u_\text{thr}$ is set to zero. 
The threshold value $u_\text{thr}$ acts as a tuning parameter to conserve control effort and minimize the on-off cycling of the RCDs, thereby reducing power consumption and prolonging actuator life. 
Figure~\ref{Fig:PWM} illustrates an example of how the continuous MPC input is modified into a PWM pulse with the inclusion of a threshold at each momentum management time step.

It is worth noting that there exists a quantization method in the literature~\cite{caverly2018off,caverly2020electric} to optimize the on-off times of a single PWM pulse in a manner that minimizes the predicted difference in system response with and without quantization. Although this approach can lead to improved performance, it comes with much great computational expense, as it involves solving a nonlinear optimization problem at each timestep. For this reason, the simplified quantization approach with thresholding described in this section is adopted in this work.


\subsection{Momentum Management Strategy 1: MPC with Single-Pulse PWM Quantized RCD Input} \label{sec:MPC_formulation}

This section makes use of the results presented in Sections~\ref{sec:Linearization} through~\ref{sec:PWM} to formulate the first proposed MPC policy for momentum management. The MPC policy is introduced first through the choices made with regards to the prediction model, state constraints, input constraints, and objective function used, followed a summary of the MPC policy in the form of an optimization problem.

\subsubsection{Prediction Model}

The discrete-time LTI prediction model outlined in Eq.~\eqref{eq:DT_dynamics_ss} is used to propagate the state dynamics at discrete time steps $j$ across the prediction horizon $N$, starting at time $t_k$. These dynamics are described by
\beq
\label{eq:MPC_Model1}
\mbf{x}_{j+1|t_k} = \mbf{A}_{k}\mbf{x}_{j|t_k} + \mbf{B}_{w,k}\mbf{w}_{j|t_k} + \mbf{B}_{u,k}^{-}\mbf{u}_{j|t_k} + \mbf{B}_{u,k}^{+}\mbf{u}_{j+1|t_k},
\eeq
where the subscript notation $j|t_k$ indicates predictions made $j$ steps ahead of the current time $t_k$. 
For example, $\mbf{x}_{j|t_k}$ is the predicted state at step $j$ from time $t_k$. The state of the model is given by $\mbf{x}_{j|t_k} = \Big[ \mbs{\theta}_{j|t_k}^\trans \,\,\, {\mbs{\omega}^{ba}_{b,j|t_k}}^\trans \,\,\, \mbf{h}^{\text{RWs}^\trans}_{b,j|t_k} \,\,\,  \mbf{e}_{j|t_k}^{\text{int}^\trans} \Big]^\trans$, and the disturbance torque is given by $\mbf{w}_{j|t_k}$. The control input vector is partitioned as $\mbf{u}_{j|t_k} = \Big[ \mbf{u}^{\text{AMT}^\trans}_{j|t_k} \,\,\, u^\text{RCD}_{j|t_k} \Big]^\trans$, 
distinguishing between the AMT inputs and the RCD input.
The matrices $\mbf{A}_{k}$, $\mbf{B}_{w,k}$, $\mbf{B}_{u,k}^{-} = \Big[ \mbf{B}_{u1,k}^{-} \,\,\, \mbf{B}_{u2,k}^{-} \Big] $, and $\mbf{B}_{u,k}^{+} = \Big[\mbf{B}_{u1,k}^{+}  \,\,\, \mbf{B}_{u2,k}^{+} \Big] $ are updated every momentum management time step (the current time step at time $t_k$), and then held as constant for the duration of the prediction horizon.

\subsubsection{State Constraints}

Inequality constraints on the system's states are imposed to ensure the practical feasibility of the controller. State constraints $\mbf{x}_{\text{min}} \leq \mbf{x}_{j|t_k} \leq \mbf{x}_{\text{max}}$  enforce bounded deviations in attitude, angular velocity, and RWs angular momentum. Specifically, these constraints include $\mbs{\theta}_\text{min} \leq \mbs{\theta}_{j|t_k} \leq \mbs{\theta}_\text{max}$, $\mbs{\omega}^{ba}_{b,\text{min}} \leq \mbs{\omega}_{b,j|t_k}^{ba} \leq \mbs{\omega}^{ba}_{b,\text{max}}$ and $\mbf{h}^\text{RWs}_{b,\text{min}} \leq \mbf{h}_{b,j|t_k}^\text{RWs} \leq \mbf{h}^\text{RWs}_{b,\text{max}}$.
Note that the desired attitude and angular velocity are chosen to be $\mbs{\theta}_d = \mbf{0}$ and $\dot{\mbs{\theta}}_{d} = \mbf{0}$ in this paper for simplicity, but they can be extended to non-zero desired trajectories based on mission requirements if necessary. The constraints on RW angular momentum can be chosen based on their saturation limits.

In order to maintain a suitable margin of operation for the RWs, it is desired to keep their angular momentum away from the saturation limits. This can be embedded into the MPC policy in the form of a soft constraint based on a threshold that is lower than the RWs' saturation limit. 
A slack variable method~\cite{de1994constraint,kerrigan2000soft} is used with the introduction of the design variable $\mbs{\alpha} \geq \mbf{0}$, such that 
\beq
    \mbf{h}_{b,\text{min}}^\text{soft} - \mbs{\alpha} \leq \mbf{h}^\text{RWs}_{b,j|t_k} \leq  \mbf{h}_{b,\text{max}}^\text{soft} + \mbs{\alpha},
\eeq
where the soft constraint limits are given by $\mbf{h}_{b,\text{min}}^\text{soft}$ and $\mbf{h}_{b,\text{max}}^\text{soft}$, and $\mbs{\alpha}$ is penalized in the MPC policy's objective function in a quadratic fashion. 
This approach does not penalize the RWs' angular momentum when $\mbf{h}_{b,\text{min}}^\text{soft} \leq \mbf{h}^\text{RWs}_{b,j|t_k} \leq  \mbf{h}_{b,\text{max}}^\text{soft}$, as in this case the constraint is satisfied with $\mbs{\alpha} = \mbf{0}$. Once these soft limits are exceeded, the soft constraint and quadratic penalty on $\mbs{\alpha}$ help keep the angular momentum away from the saturation limit.
This allows MPC to preemptively react to momentum buildup before RW saturation, and relaxes the potential feasibility issue of using the saturation limit as a hard constraint. The values of $\mbf{h}^\text{soft}_{b,\text{min}}$ and $\mbf{h}^\text{soft}_{b,\text{max}}$ are selected based on the operation scenario and desired safety margin.

\subsubsection{Input Constraints}

Actuator constraints are implemented to limit the input magnitude and AMT translational rates through the inequalities $\mbf{u}_{\text{min}} \leq \mbf{u}_{j|t_k} \leq \mbf{u}_{\text{max}}$ and $\dot{\mbf{u}}^\text{AMT}_{\text{min}} \leq \dot{\mbf{u}}^\text{AMT}_{j|t_k} \leq \dot{\mbf{u}}^\text{AMT}_{\text{max}}$. Due to the discrete-time nature of the control inputs, the rate constraint is implemented as
\beq
\dot{\mbf{u}}^\text{AMT}_{\text{min}} \leq \frac{\mbf{u}^\text{AMT}_{j+1|t_k} - \mbf{u}^\text{AMT}_{j|t_k}}{t_{k+1} - t_k} \leq \dot{\mbf{u}}^\text{AMT}_{\text{max}}.
\eeq

To ensure continuity in the FOH discretization of the AMT input, the initial $j = 0$ AMT control input at each time step is matched to the $j = 1$ AMT input of the previous momentum management timestep, which is described through the constraint $\mbf{u}^{\text{AMT}}_{0|t_k} = \mbf{u}^{\text{AMT}}_{1|t_{k-1}}$.

\subsubsection{Objective Function}

The objective function of the proposed MPC policy is chosen as
\begin{multline}
\sum_{j=0}^{N-1} \Bigl{(} \mbf{x}_{j|t_k}^\trans \mbf{Q} \mbf{x}_{j|t_k} + \mbf{u}_{j|t_k}^\trans \mbf{R} \mbf{u}_{j|t_k} + \tilde{\mbf{u}}^{\text{AMT}^\trans}_{j|t_k} \tilde{\mbf{R}}\tilde{\mbf{u}}^\text{AMT}_{j|t_k}\Bigl{)} \\ + \mbf{x}_{N|t_k}^\trans \mbf{P} \mbf{x}_{N|t_k} + \mbf{u}_{N|t_k}^\trans \mbf{R}_N \mbf{u}_{N|t_k} + \mbs{\alpha}^\trans\mbf{C}\mbs{\alpha}
\end{multline}
where the weights $\mbf{Q} = \mbf{Q}^\trans$ and $\mbf{R} = \mbf{R}^\trans$ are positive semi-definite and positive definite matrices, respectively, penalizing the state and control input over the prediction horizon of $N$ time steps.
The terminal costs, $\mbf{P}$ and $\mbf{R}_N$, are associated with penalties on the state and input at the final step. The weight $\mbf{C} = \mbf{C}^\trans$ is a positive semi-definite matrix used to adjust the emphasis on maintaining the soft constraint. 
The penalty $\mbs{\alpha}^\trans\mbf{C}\mbs{\alpha}$ is activated only when soft constraint violation ($\mbs{\alpha} > \mbf{0}$). A higher value for $\mbf{C}$ places a stronger priority on minimizing the violation.
The term $\tilde{\mbf{u}}^{\text{AMT}^\trans}_{j|t_k} \tilde{\mbf{R}}\tilde{\mbf{u}}^\text{AMT}_{j|t_k}$ within the objective function is defined with the variable $\tilde{\mbf{u}}^\text{AMT}_{j|t_k} = \mbf{u}^\text{AMT}_{j+1|t_k}-\mbf{u}^\text{AMT}_{j|t_k}$ and the positive definite weighting matrix $\mbftilde{R} = \mbftilde{R}^\trans$ is included to penalize translation of the AMT. This helps maintain the AMT at a stationary location, rather than allowing it to be in constant motion.

\subsubsection{Summary of MPC Policy for Momentum Management Strategy~1}
\label{sec:Strategy1Summary}

The proposed MPC policy for Momentum Management Strategy~1 involves solving the optimization problem
\begin{align}
    &\minimize_{\mbs{\mathcal{X}},\hspace{2pt} \mbs{\mathcal{U}},\hspace{2pt} \mbs{\alpha}} \sum_{j=0}^{N-1} \Bigl{(} \mbf{x}_{j|t_k}^\trans \mbf{Q} \mbf{x}_{j|t_k} + \mbf{u}_{j|t_k}^\trans \mbf{R} \mbf{u}_{j|t_k} + \tilde{\mbf{u}}^{\text{AMT}^\trans}_{j|t_k} \tilde{\mbf{R}}\tilde{\mbf{u}}^\text{AMT}_{j|t_k}\Bigl{)} \nonumber \\ &\hspace{15em} + \mbf{x}_{N|t_k}^\trans \mbf{P} \mbf{x}_{N|t_k} + \mbf{u}_{N|t_k}^\trans \mbf{R}_N \mbf{u}_{N|t_k} + \mbs{\alpha}^\trans\mbf{C}\mbs{\alpha} \label{eq:MPC_cost2}\\
    &\hspace{0em}\text{subject to } \nonumber\\
    &\hspace{2em} \mbf{x}_{j+1|t_k} = \mbf{A}_{k}\mbf{x}_{j|t_k} + \mbf{B}_{w,k}\mbf{w}_{j|t_k} + \mbf{B}_{u,k}^{-}\mbf{u}_{j|t_k} + \mbf{B}_{u,k}^{+}\mbf{u}_{j+1|t_k} , \,\, j=0,\cdots,N-1,  \nonumber\\
    &\hspace{2em} \mbf{x}_{0|t_k} = \mbf{x}(t_k), \nonumber\\
    &\hspace{2em} \mbf{u}^\text{AMT}_{0|t_k} = \mbf{u}^\text{AMT}_{1|t_{k-1}}, \nonumber\\
    &\hspace{2em} \mbf{x}_{\text{min}} \leq \mbf{x}_{j|t_k} \leq \mbf{x}_{\text{max}}, \hspace{1em} j=0,\cdots,N, \nonumber\\
    &\hspace{2em} \mbf{u}_{\text{min}} \leq \mbf{u}_{j|t_k} \leq \mbf{u}_{\text{max}}, \hspace{1em} j=0,\cdots,N, \nonumber\\
    &\hspace{2em} \dot{\mbf{u}}^\text{AMT}_{\text{min}} \leq \frac{\mbf{u}^\text{AMT}_{j+1|t_k} - \mbf{u}^\text{AMT}_{j|t_k}}{t_{k+1} - t_k} \leq \dot{\mbf{u}}^\text{AMT}_{\text{max}}, \hspace{1em} j=0,\cdots,N-1, \nonumber\\
    &\hspace{2em} \mbf{h}_{b,\text{min}}^\text{soft} - \mbs{\alpha} \leq \mbf{h}^\text{RWs}_{b,j|t_k} \leq  \mbf{h}_{b,\text{max}}^\text{soft} + \mbs{\alpha}, \hspace{1em} j=0,\cdots,N, \nonumber\\
    &\hspace{2em} \mbs{\alpha} \geq \mbf{0}, \nonumber
\end{align}
where $\mbs{\mathcal{X}} = \mbf{x}_{0|t_k},\mbf{x}_{1|t_k},\cdots, \mbf{x}_{N|t_k}$, $\mbs{\mathcal{U}} = \mbf{u}_{0|t_k},\mbf{u}_{1|t_k},\cdots, \mbf{u}_{N|t_k}$, $ \mbs{\alpha} \in \mathbb{R}^{3\times 1}$ are the design variables, $N$ is the number of timesteps in the prediction horizon, and $\mbf{x}(t_k)$ is the known system state at time $t_k$. 
Note that it is common for MPC policies to optimize the input sequence $\mbs{\mathcal{U}} = \mbf{u}_{0|t_k},\mbf{u}_{1|t_k},\cdots, \mbf{u}_{N-1|t_k}$, however, the input at the end of the prediction horizon, $\mbf{u}_{N|t_k}$, must be considered here to account for the FOH discretization of the AMT translation.
This optimization problem can be solved as a QP using well-established solvers. This QP can be solved efficiently and accurately, as it is a convex optimization problem. 
The MATLAB function \texttt{quadprog} is used to solve the QP optimization problem in the simulation results of Section~\ref{sec:Num_Sim}. 

Once this optimization problem is solved, the optimal control inputs associated with the first two time steps, $\mbf{u}_{0|t_k} = \Big[\mbf{u}^{\text{AMT}^\trans}_{0|t_k} \,\,\, u^\text{RCD}_{0|t_k} \Big]^\trans$ and $\mbf{u}_{1|t_k} = \Big[\mbf{u}^{\text{AMT}^\trans}_{1|t_k} \,\,\, u^\text{RCD}_{1|t_k} \Big]^\trans$, are used to determine the AMT and RCD inputs over the time interval $t_k \leq t \leq t_{k+1}$. Specifically, the AMT position over this time interval is chosen as the linear interpolation of $\mbf{u}^{\text{AMT}}_{0|t_k}$ and $\mbf{u}^{\text{AMT}}_{1|t_k}$, as designed through the use of a FOH discretization. The RCD on-off times during this time interval are found by using the single-pulse PWM quantization approach outlined in Section~\ref{sec:PWM} on the value $u^\text{RCD}_{0|t_k} $. The control inputs are made over the time interval $t_k \leq t \leq t_{k+1}$ and then this entire process is performed again at timestep $t_{k+1}$.


\subsection{Momentum Management Strategy 2: Iterative Backwards-in-Time MPC with Single-Pulse PWM Quantized RCD Input} \label{sec:iter_back}

The MPC policy in Section~\ref{sec:MPC_formulation} provides a promising momentum management controller that can be implemented on a solar sail system.
However, since the PWM-inspired quantization method discussed in Section~\ref{sec:PWM} alters the optimal RCD input into a single pulse suitable for practical implementation, the MPC algorithm does not inherently account for this quantization when determining this optimal input, which affects its predictive capabilities.
Incorporating knowledge of this quantization into the MPC framework has the potential to improve overall performance by aligning the prediction model closer to its practical implementation. 
The challenge with this is that the quantized RCD input is an integer variable, which transforms the MPC optimization problem into a mixed-integer optimization problem that cannot be solved efficiently and reliably. Typically, mixed-integer optimization problems require computationally-expensive branch-and-bound solution techniques~\cite[Chapter~8.3]{martins2021engineering}.

To address this, a framework to iteratively solve the mixed-integer MPC optimization problem as a sequence of QPs is proposed in this section.
This iterative method quantizes the RCD input at the end of the prediction horizon step using a single PWM on-off pulse. The MPC optimization problem is then re-solved, treating this quantized input as a fixed virtual control input. With each iteration of the algorithm, the RCD input at the end of the shrinking horizon is fixed as a PWM-quantized input, while the remaining variables in the prediction horizon are optimized. Notably, only the RCD inputs are fixed during this process; state variables and AMT inputs remain as free design variables. The remainder of this section formulates the equations needed for this proposed iterative quantization strategy, followed by a summary of the MPC policy.

\subsubsection{Formulation of Iterative Backwards-in-Time MPC with Quantization}

Consider a single momentum management time step $t_k \leq t \leq t_{k+1}$, where the quantized PWM RCD on-off pulse is initiated at $t=t_k$ and cut-off at $t = t_k + t_c$, where $t_c = \Delta t \cdot \f{u^\text{RCD}_{k,\text{mpc}}}{u^\text{RCD}_\text{max}}$ is the pulse duration within a single timestep $\Delta t$, and $u_\text{mpc}$ is the MPC optimal continuous RCD input value as defined in Section~\ref{sec:PWM}.
This momentum management timestep can be divided into two parts, one with a fixed $u^\text{RCD}_\text{max}$ input for $t_k \leq t \leq t_k + t_c$, and the rest without any RCD input for $t_k + t_c < t \leq t_{k+1}$. Assuming that $\mbf{A}_{\ell} = \mbf{A}(t_k)$, $\mbf{B}_{w,\ell} = \mbf{B}_w(t_k)$, and $\mbf{B}_{u,\ell} = \Big[\mbf{B}_{u1}(t_k) \,\,\, \mbf{B}_{u2}(t_k) \Big] $ are constant across the prediction horizon, the response of the state is given by
\begin{multline}
    \mbf{x}_{k+1}     = e^{\mbf{A}_\ell(t_{k+1} - t_k)} \mbf{x}_k + \int^{t_{k+1}}_{t_k} e^{\mbf{A}_\ell(t_{k+1}-\tau)}\mbf{B}_w(t_k) \mbf{w}(\tau)\dee\tau \\ \\ + \int^{t_{k+1}}_{t_k} e^{\mbf{A}_\ell(t_{k+1}-\tau)}\mbf{B}_{u1}(t_k) \mbf{u}^\text{AMT}(\tau)\dee\tau 
    + \int^{t_{k+1}}_{t_k} e^{\mbf{A}_\ell(t_{k+1}-\tau)}\mbf{B}_{u2}(t_k) u^\text{RCD}(\tau)\dee\tau \label{eq:MPC2_StateEquation2}
\end{multline}
Taking advantage of the ZOH discretization on $\mbf{w}$, the FOH discretization on $\mbf{u}^\text{AMT}$, as well as the ZOH discretization on $u^\text{RCD}$ from $t_k \leq t \leq t_k + t_c$ and the lack of any RCD input from $t_k + t_c < t \leq t_{k+1}$,~\eqref{eq:MPC2_StateEquation2} can be rewritten as
\beq
    \mbf{x}_{k+1}  = \mbf{A}_{k} \mbf{x}_k + \mbf{B}_{w,k}^{-} \mbf{w}_{k} + \bbm \mbf{B}_{u1,k}^{-} & \mbf{B}_{u1,k}^{+} \ebm \bbm \mbf{u}^\text{AMT}_{k} \\\mbf{u}^\text{AMT}_{k+1} \ebm + \mbf{A}_{c} \mbf{B}_{u2,c}^{-} u^\text{RCD}_\text{max},
\eeq
where $\mbf{A}_{c} = e^{\mbf{A}_\ell (t_{k+1} -t_c-t_k)}$, $\mbf{B}_{u2,c}^{-} = \int^{t_k + t_c}_{t_k} e^{\mbf{A}_\ell(t_k+t_{c}-\tau)} \mbf{B}_\ell \mbf{u}(\tau)\dee\tau$ are the ZOH Jacobian matrices evaluated at time $t_k + t_c$, and $u^\text{RCD}_\text{max}$ is a constant torque when the RCD is turned on.
This decomposition propagates the state with PWM-quantized RCD torque where $u^\text{RCD}(\tau) = u^\text{RCD}_\text{max}$ when $\tau \in [t_k, t_k+t_c]$, and $u^\text{RCD}(\tau) = 0$ when $\tau \in (t_k+t_c, t_{k+1}]$, while $\mbf{A}_{k}$, $\mbf{B}_{w,k}^{-} \mbf{w}_{k}$, $\mbf{B}_{u1,k}^{-}\mbf{u}^\text{AMT}_{k}$, $\mbf{B}_{u1,k}^{+}\mbf{u}^\text{AMT}_{k+1}$ remain the same as in~\eqref{eq:DT_dynamics_ss}.

In order to iteratively solve the MPC optimization problem by fixing the quantized RCD input at the end of the prediction horizon, the prediction model in~\eqref{eq:MPC_Model1} is replaced with
\beq
    \mbf{x}_{j+1|t_k} = \mbf{A}_{k}\mbf{x}_{j|t_k} + \mbf{B}_{w,k}^{-} \mbf{w}_{j|t_k} + \bbm \mbf{B}_{u1,k}^{-} & \mbf{B}_{u1,k}^{+} \ebm \bbm \mbf{u}^\text{AMT}_{j|t_k} \\\mbf{u}^\text{AMT}_{j+1|t_k} \ebm + \mbf{B}_{u2,k}^{-}u^\text{RCD}_{j|t_k},
\eeq
    for $j = 0, 1, \cdots, N-N_{bk}-1$ and
\beq
    \mbf{x}_{j+1|t_k} = \mbf{A}_{k}\mbf{x}_{j|t_k} + \mbf{B}_{w,k}^{-} \mbf{w}_{j|t_k} + \bbm \mbf{B}_{u1,k}^{-} & \mbf{B}_{u1,k}^{+} \ebm \bbm \mbf{u}^\text{AMT}_{j|t_k} \\\mbf{u}^\text{AMT}_{j+1|t_k} \ebm + \mbf{A}_{c}\mbf{B}_{u2,c}^{-}u^\text{RCD}_\text{max},
\eeq
for $j = N-N_{bk}, \cdots, N-1$, 
where $N_{bk} = 1, 2, \cdots, N-1$ is the number of iterations performed since the initial MPC solution and is also the number of timesteps at the end of the horizon with a quantized RCD input.

\begin{figure}[h!]
    \centering
        \includegraphics[width=0.97\textwidth]{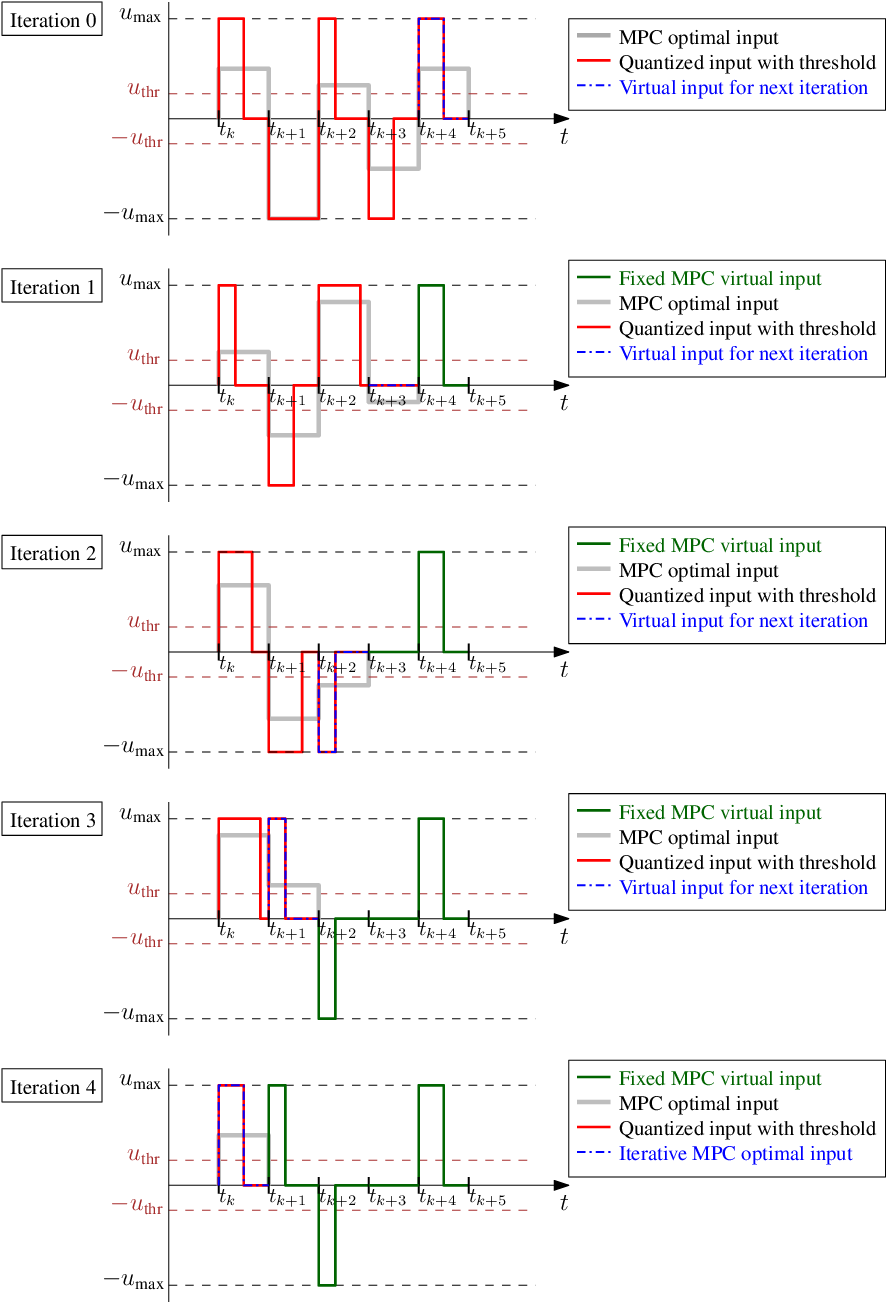}
    \caption{Iterative solving MPC backwards in time with a prediction horizon of $N=5$ and quantized inputs with the PWM-inspired method. }\label{Fig:iter_back}
\end{figure}

The proposed iterative backwards-in-time MPC policy operates by performing a sequential PWM quantization to the furthest optimal RCD input in the prediction horizon, and iteratively re-optimize the remainder of the unmodified sequence with $N_{bk}$ fixed PWM-quantized RCD inputs at the end of the prediction horizon.
Note that every iteration of the iterative backwards MPC algorithm removes one more time step of RCD input from the end of the prediction horizon with a fixed PWM-based quantization value of $\mbf{A}_{c} \mbf{B}_{u2,c}^{-} u^\text{RCD}_\text{max}$. 
In the first iteration, only the RCD input from the final time step in the prediction horizon is fixed as a PWM on-off pulse, while the AMT inputs remain free for optimization. In subsequent iterations, one additional RCD input is quantized and fixed, progressing step by step closer to the current time step.
The modified dynamics are incorporated into the MPC policy as equality constraints, replacing the continuous-time model with the PWM-based quantized inputs as appropriate. The iterative process continues $N_{bk}$ times until $N_{bk} = N-1$, at which point all RCD inputs in the prediction horizon are quantized other than the first time step. Finally, the first time step’s input is extracted and implemented as a PWM-quantized input, as outlined in Section~\ref{sec:PWM}. This method ensures that the MPC solution incorporates knowledge of PWM quantization, improving dynamic model fidelity in the MPC prediction. Figure~\ref{Fig:iter_back} illustrates the sequential iterative backward MPC algorithm, showing how the far-end RCD input is fixed at each iteration and the remaining design variables are re-optimized.
Since fixed quantized RCD inputs no longer contribute to the cost function, their influence is gradually removed during iterations of the backwards iteration procedure. To maintain consistent weighting in the cost function, a scaling factor of $N/(N-N_{bk})$ is applied to the remaining RCD input design variables. This ensures that the optimization problem remains consistent across iterations, yielding an optimal solution that respects the quantization constraints.

\subsubsection{Summary of MPC Policy for Momentum Management Strategy~2}

The proposed MPC policy for Momentum Management Strategy~2 involves iteratively solving the optimization problem
\begin{align}
    &\minimize_{\mbs{\mathcal{X}},\hspace{2pt} \mbs{\mathcal{U}},\hspace{2pt} \mbs{\alpha}} \sum_{j=0}^{N-1} \bigg{(} \mbf{x}_{j|t_k}^\trans \mbf{Q} \mbf{x}_{j|t_k} + \mbf{u}_{j|t_k}^{\text{AMT}^\trans} \mbf{R} \mbf{u}_{j|t_k}^\text{AMT} + \tilde{\mbf{u}}^{\text{AMT}^\trans}_{j|t_k} \tilde{\mbf{R}}\tilde{\mbf{u}}^\text{AMT}_{j|t_k} \nonumber \\ &\hspace{2em}+ \f{N}{N-N_{bk}}\mbf{u}_{j|t_k}^{\text{RCD}^\trans} \mbf{R} \mbf{u}_{j|t_k}^\text{RCD}\bigg{)} 
    + \mbf{x}_{N|t_k}^\trans \mbf{P} \mbf{x}_{N|t_k} + \mbf{u}_{N|t_k}^\trans \mbf{R}_N \mbf{u}_{N|t_k} + \mbs{\alpha}^\trans\mbf{C}\mbs{\alpha} \label{eq:MPC_cost_iter}\\
    &\hspace{0em}\text{subject to } \nonumber\\
    &\hspace{2em} \mbf{x}_{j+1|t_k} = 
    \begin{cases} 
    &\mbf{A}_{k}\mbf{x}_{j|t_k} + \mbf{B}_{w,k}^{-} \mbf{w}_{j|t_k} + \bbm \mbf{B}_{u1,k}^{-} & \mbf{B}_{u1,k}^{+} \ebm \bbm \mbf{u}^\text{AMT}_{j|t_k} \\\mbf{u}^\text{AMT}_{j+1|t_k} \ebm + \mbf{B}_{u2,k}^{-}u^\text{RCD}_{j|t_k}, \nonumber \\ &\hspace{12em} \text{for} \hspace{0.5em} j = 0, 1, \cdots, N-N_{bk}-1 ,\nonumber\\
    &\mbf{A}_{k}\mbf{x}_{j|t_k} + \mbf{B}_{w,k}^{-} \mbf{w}_{j|t_k} + \bbm \mbf{B}_{u1,k}^{-} & \mbf{B}_{u1,k}^{+} \ebm \bbm \mbf{u}^\text{AMT}_{j|t_k} \\\mbf{u}^\text{AMT}_{j+1|t_k} \ebm + \mbf{A}_{c} \mbf{B}_{u2,c}^{-} u^\text{RCD}_\text{max}, \nonumber \\ &\hspace{12em} \text{for} \hspace{0.5em} j = N-N_{bk}, \cdots, N-1 , 
    \end{cases}\nonumber\\
    &\hspace{2em} u^\text{RCD}_{j|t_k} = u^\text{RCD}_\text{max} , \hspace{1em} j = N-N_{bk}, \cdots, N-1, \nonumber\\
    &\hspace{2em} \mbf{x}_{0|t_k} = \mbf{x}(t_k), \nonumber\\
    &\hspace{2em} \mbf{u}^\text{AMT}_{0|t_k} = \mbf{u}^\text{AMT}_{1|t_k-1}, \nonumber\\
    &\hspace{2em} \mbf{x}_{\text{min}} \leq \mbf{x}_{j|t_k} \leq \mbf{x}_{\text{max}}, \hspace{1em} j=0,1,\cdots,N, \nonumber\\
    &\hspace{2em} \mbf{u}_{\text{min}} \leq \mbf{u}_{j|t_k} \leq \mbf{u}_{\text{max}}, \hspace{1em} j=0,1,\cdots,N, \nonumber\\
    &\hspace{2em} \dot{\mbf{u}}^\text{AMT}_{\text{min}} \leq \frac{\mbf{u}^\text{AMT}_{j+1|t_k} - \mbf{u}^\text{AMT}_{j|t_k}}{t_{k+1} - t_k} \leq \dot{\mbf{u}}^\text{AMT}_{\text{max}}, \hspace{1em} j=0,1,\cdots,N-1, \nonumber\\
    &\hspace{2em} \mbf{h}_{b,\text{min}}^\text{soft} - \mbs{\alpha} \leq \mbf{h}^\text{RWs}_{b,j|t_k} \leq  \mbf{h}_{b,\text{max}}^\text{soft} + \mbs{\alpha}, \hspace{1em} j=0,1,\cdots,N, \nonumber\\
    &\hspace{2em} \mbs{\alpha} \geq \mbf{0}. \nonumber
\end{align}

This momentum management strategy involves solving the optimization problem described in~\eqref{eq:MPC_cost2} and the iteratively solving the optimization problem in equation~\eqref{eq:MPC_cost_iter} for $N_{bk} = 1, \cdots, N-1$. The optimal input solution from the final iteration ($N_{bk} = N-1$) is then applied to the system using the same approach outlined in Section~\ref{sec:Strategy1Summary}.
%


\section{Numerical Simulation Results} \label{sec:Num_Sim}

This section presents a numerical simulation of a solar sail system employing the proposed MPC-based momentum management techniques. The simulation parameters are chosen to reflect a realistic, large-scale solar sail system, leveraging publicly-available data from NASA’s Solar Cruiser~\cite{JohnsonLes2020SCTM, Tyler2024, johnson2019solar, Inness2023}. This setup demonstrates the feasibility and practicality of the proposed method in addressing the challenges of solar sail momentum management.

A numerical simulation demonstrating the tracking performance of RWs PID control law is presented in Section~\ref{sec:Sim_PID}, where the PID gains are tuned to track the desired attitude under a non-equilibrium initial attitude without any momentum management actuation.
The PID attitude control law successively tracks the spacecraft attitude while the RWs angular momentum keeps growing under the consistent affect of disturbances. Section~\ref{sec:Sim_NASA} presents results with the state-of-the-art momentum management strategy used on Solar Cruiser. 
Section~\ref{sec:Sim_MPC} shows the simulation results of the proposed MPC momentum management strategies formulated in Sections~\ref{sec:MPC_formulation} and~\ref{sec:iter_back}.

\subsection{Problem Setup} \label{sec:Num_Setup}

The specifications of the solar sail and its operational limits are summarized in Table~\ref{tab:table1}, based on publicly available resources for the Solar Cruiser ~\cite{JohnsonLes2020SCTM,heaton2023reflectivity,Inness2023}.
The sail is modeled as a square with dimensions of $41.72$~m $\times$ $41.72$~m $\times$ $0.001$~m, derived from the length of each boom being $29.5$~m~\cite{JohnsonLes2020SCTM}.
Attitude control is achieved using commercially-available RWs with a radius of $0.11$~m and a maximum angular momentum capacity of $1$~N$\cdot$m$\cdot$s~\cite{BCTRW}. 
As a proof of concept, the CM of the solar sail membrane, bus, and RWs are assumed to be coincident, which simplifies the dynamic model. Neglecting the height of the bus relative to the boom length minimizes coupling effects between the three axes, facilitating control system tuning, although this restriction can be accounted for in the model if desired. 

The simulation assumes a constant environmental force and torque as representative of a worst-case disturbance scenario for the Solar Cruiser under a $\ang{17}$ sun incidence angle~\cite{Inness2023}.  The SRP force and disturbance torque are set to $\mbf{f}^\text{SRP}_b = \Big[ 0.0003 \,\,\, 0 \,\,\, 0.013\Big]^\trans$~N and $\mbs{\tau}^\text{dist}_b = \Big[ 8 \,\,\, 8 \,\,\, 0.2\Big]^\trans \times 10^{-4}$~N$\cdot$m. The SRP force is computed assuming Solar Cruiser is at $1$~au and a $\ang{0}$ clock angle, and has the reflectivity properties outlined in~\cite{heaton2015update}, while the disturbance torque is chosen based on the worst-case deformed sail shapes investigated in~\cite{gauvain2023solar}. While a disturbance of such a magnitude is unlikely to persist throughout a practical mission, it provides a conservative estimate for validating the robustness of the proposed controller.

The momentum management actuators are modeled with realistic operational limits.
The AMT has a translation limit of $\mbf{u}^\text{AMT}_{\text{max}} = \Big[ 0.29 \,\,\, 0.29\Big]^\trans$~m and a rate limit of $\dot{\mbf{u}}^\text{AMT}_{\text{max}} =  \Big[ 0.5 \,\,\, 0.5\Big]^\trans$~mm/s for both $+/-$ directions in the $\vect{b}^1$ and $\vect{b}^2$ axes~\cite{JohnsonLes2020SCTM}.
 The roll torque generated when the RCDs are turned on is set to meet the Solar Cruiser’s roll torque requirement at $6.525\times 10^{-5}$~N$\cdot$m, which is $1.5$ times the sum of worst case roll disturbance and AMT induced roll torque at its maximum position offset~\cite{heaton2023reflectivity,johnson2022nasa}.
Given that the RCD torque magnitude is only $1.5$ times larger than the roll-axis disturbance/AMT-coupled torque, the RCDs must activate frequently to prevent angular momentum buildup. Steady-state operation is expected to exhibit RCD activation approximately two-thirds of the time, with the remainder spent in an ``off'' state. 

The simulation utilizes a discrete time step of $\dee t = 1$~sec for RW attitude control, while the momentum management time step is set to $\Delta t = 100$~sec.
The tuning parameters of the MPC policy, including prediction horizon ($N$), weighting matrices ($\mbf{Q}$, $\mbf{R}$, $\tilde{\mbf{R}}$, $\mbf{C}$), terminal costs ($\mbf{P}$, $\mbf{R}_N$), soft constraints ($\mbf{h}_{b,\text{min}}^\text{soft}$, $\mbf{h}_{b,\text{max}}^\text{soft}$), and RCD thresholds are discussed further in Section~\ref{sec:Sim_MPC}.

\begin{table}[t!]
\caption{\label{tab:table1} Solar sail specification used in numerical simulation}
\centering
\begin{tabular}{lccc}
Parameter & Value & Unit \\
\hline
bus mass, $m_p$ & $50$ & kg \\
sail mass, $m_s$ & $44.6$ & kg \\
bus dimension, $\ell_1\times \ell_2\times \ell_3$ & $0.9\times0.9\times0.3$ & $\text{m}\times\text{m}\times\text{m}$ \\
sail dimension, $L\times L\times h_s$ & $41.72\times41.72\times0.001$ & $\text{m}\times\text{m}\times\text{m}$ \\
nominal inertia of bus, $ \mbf{J}^{\mathcal{P}p}_b$ & $\textrm{diag}(3.75, 3.75, 6.75)$ & kg$\cdot$m$^2$\\
nominal inertia of sail, $\mbf{J}^{\mathcal{S}s}_b$ & $\textrm{diag}(6468.9, 6468.9, 12937.7)$ & kg$\cdot$m$^2$\\
RW radius, $R$& $0.11$ & m \\
RW angular momentum capacity 
& $1$ & N$\cdot$m$\cdot$s\\
AMT range, $\Big[ r^{\text{AMT}}_{b1,\text{max}} \,\,\, r^{\text{AMT}}_{b2,\text{max}} \Big]^\trans $ & $\big[ \pm 0.29 \,\,\, \pm 0.29\big]^\trans$ & m\\
AMT speed, $\Big[ \dot{r}^{\text{AMT}}_{b1,\text{max}} \,\,\, \dot{r}^{\text{AMT}}_{b2,\text{max}} \Big]^\trans $ & $ \big[ \pm 0.5 \,\,\, \pm 0.5\big]^\trans$ & mm/s\\
RCD torque, $u^\text{RCD}_\text{max} = \tau_{b3,\text{on}}^\text{RCD}$ & $6.525\times 10^{-5}$  & N$\cdot$m\\
constant SRP force, $\mbf{f}^\text{SRP}_b$ & $\big[ 0.0003 \,\,\, 0 \,\,\, 0.013\big] ^\trans$ &N \\
disturbance torque, $\mbf{w} = \mbs{\tau}^\text{dist}_b$ & $\big[ 8 \,\,\, 8 \,\,\, 0.2\big] ^\trans \times 10^{-4}$ & N$\cdot$m \\
\hline
\end{tabular}
\end{table}                                                                        


\subsection{Attitude Tracking with RW PID Control} \label{sec:Sim_PID}

This section presents the simulation setup and the solar sail's attitude tracking performance using a RW PID controller. The desired attitude maneuver involves a $\ang{2}$ slew in the yaw axes and a $\ang{1}$ slew in the roll axis, returning to an attitude with $\mbs{\theta}_d=\mbf{0}$. In this scenario, the RWs strive to achieve the desired attitude while countering constant external disturbances. 
The initial conditions of the simulation are $\mbs{\theta}_0 = \Big[ 2 \,\,\, 0 \,\,\, 1\Big]^\trans$~deg, $\mbs{\omega}^{ba}_{b,0} = \mbf{0}$~deg/s, $\mbf{h}^{\text{RWs}}_{b,0} = \mbf{0}$~N$\cdot$m/s, and $\mbf{e}^\text{int}_0 = \mbf{0}$~deg$\cdot$s, where $\mbf{e}^\text{int} = \int^t_{t_0}( \mbs{\theta}(\tau) -  \mbs{\theta}_d)\dee\tau$ is the internal state representing the integral term of PID law, such that $\delta \dot{\mbf{e}}^\text{int} =  \delta\mbs{\theta}$.
The PID gains are chosen as $\mbf{K}_p = 0.4\cdot\mbf{1}_{3\times 3}$~N$\cdot$m/rad, $\mbf{K}_d = 140 \cdot\mbf{1}_{3\times 3}$~N$\cdot$m$\cdot$s/rad, and $\mbf{K}_i = 10^{-3}\cdot \mbf{1}_{3\times 3}$~N$\cdot$m/(rad$\cdot$s).
The constant disturbance torque $\mbs{\tau}^\text{ext}_b$ is set to the maximum expected disturbance for NASA’s Solar Cruiser, as defined in Section~\ref{sec:Num_Setup}.

\begin{figure}[!t]
\centering
\subfigure[attitude]
{
		\includegraphics[width=0.65\textwidth]{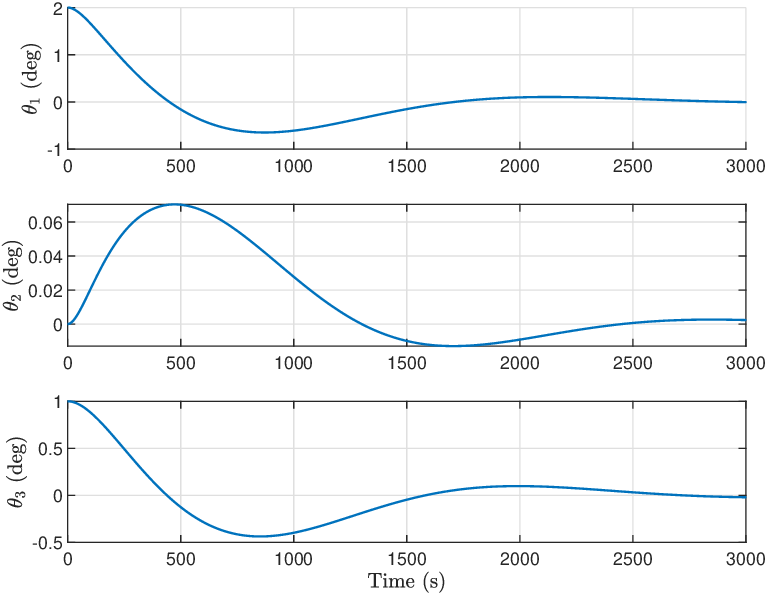} %
		\label{Fig_RWpid_1}
}
\subfigure[RWs angular momentum]
{
		\includegraphics[width=0.65\textwidth]{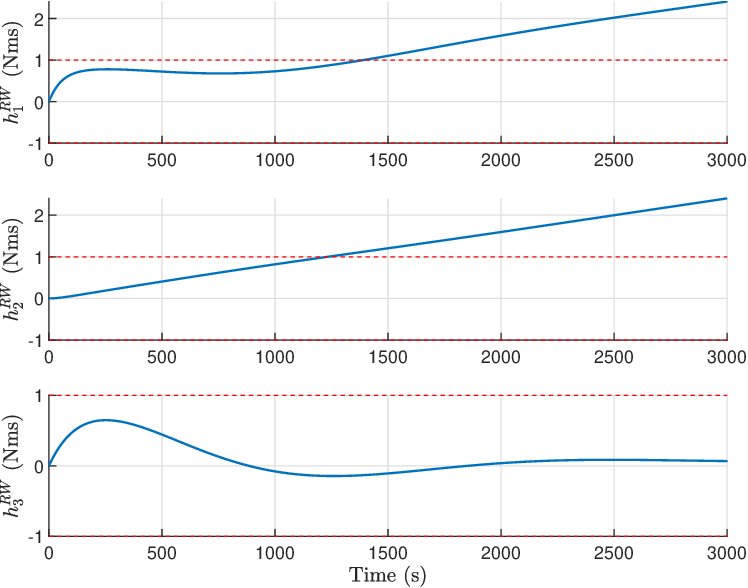} %
		\label{Fig_RWpid_2}
}
\centering
\vspace{-8pt}
\caption{Simulation results of the spacecraft attitude and RWs angular momentum using a PID attitude control law without any momentum management control. The red dashed lines in Figure~\ref{Fig_RWpid_2} indicate the saturation limit of RWs.}\label{Fig_RWpid}
\end{figure}

Numerical simulation are carried using the nonlinear solar sail attitude dynamics in~\eqref{EOM_2}. Given that the maximum translation rate of the AMT is $\pm 0.5$~mm/s for Solar Cruiser~\cite{JohnsonLes2020SCTM}, it is assumed that $\ddot{\mbf{r}}_b^{ps} = \mbf{0}$ in the simulation, and a first-order hold on the AMT input allows for the substitution $\dot{\mbf{r}}_b^{ps}(t_k) = \f{\mbf{r}_b^{ps}(t_{k+1}) - \mbf{r}_b^{ps}(t_k)}{t_{k+1} - t_k}$ between time steps $[t_k, t_{k+1})$.
The simulation results for the slew over $3000$~seconds without any momentum management is shown in Figure~\ref{Fig_RWpid}.
Figure~\ref{Fig_RWpid_1} illustrates the attitude tracking performance, while Figure~\ref{Fig_RWpid_2} depicts the evolution of angular momentum in the RWs.
As the RWs adjust the solar sail's attitude from $\mbs{\theta}_0 = \Big[ 2 \,\,\, 0 \,\,\, 1\Big]^\trans$~deg to $\mbs{\theta}_d=\mbf{0}$~deg and maintain the desired orientation, they simultaneously counteract the disturbance torque, leading to a steady accumulation of angular momentum. This buildup continues unabated, as shown in Figure~\ref{Fig_RWpid_2}, where the RW angular momentum exceeds its maximum capacity (indicated by the red dashed line). In practice, this would result in RW saturation, rendering them ineffective for further attitude control.
Although the disturbance in the roll-axis ($\vect{b}^3$ axis) is relatively small, the angular momentum still exhibits a consistent growth trend at steady state after the slew maneuver, underscoring the cumulative nature of angular momentum under constant disturbances.

These results emphasize the critical need for momentum management to preserve RW control authority over the operational lifespan of the solar sail. The subsequent sections detail the performance of Solar Cruiser's state-of-the-art PID-based momentum management algorithm, as well as the proposed MPC-based optimal momentum management framework, demonstrating the ability to mitigate RW saturation and maintain system operability.


\subsection{Solar Cruiser's PID Momentum Management} \label{sec:Sim_NASA}

As a comparison to the state of the art, this section presents a momentum management strategy that is planned for Solar Cruiser's AMT and RCDs actuation based on~\cite{Inness2023,Tyler2024}.
Solar Cruiser's PID-based momentum management strategy utilizes predefined on-off thresholds for RW angular momentum in the pitch/yaw and roll axes to govern AMT and RCD activation. 
Specifically, an actuator is engaged once the stored RW momentum exceeds an upper activation threshold and remains active until the momentum decreases below a lower deactivation threshold.
This approach prevents unnecessary rapid switching (chattering) and ensures that AMT and RCDs operate only when needed.

A set of on-off threshold of RWs stored angular momentum in pitch/yaw and roll axes is chosen for AMT and RCDs activation and deactivation commands.
The activation threshold is designed higher than the deactivation threshold, such that a momentum management actuator is activated once the angular momentum of the corresponding RW axis exceeds the activation threshold, and is deactivated until the RW angular momentum decreases to the value below of deactivate threshold.

The RCDs' actuation follows a simple on-off logic with a fixed torque magnitude when activated. 
Conversely, the AMT translation employs two decoupled PID control laws based on the stored RW momentum, where
\begin{align}
    u^{\text{AMT}}_{1} &= K^\text{AMT}_p h^\text{RWs}_{b2} +K^\text{AMT}_d \dot{h}^\text{RWs}_{b2} +K^\text{AMT}_i \int^t_{t_0} h^\text{RWs}_{b2}(\tau) \dee\tau,\\
    u^{\text{AMT}}_{2} &= -K^\text{AMT}_p h^\text{RWs}_{b1} -K^\text{AMT}_d \dot{h}^\text{RWs}_{b1} -K^\text{AMT}_i \int^t_{t_0} h^\text{RWs}_{b1}(\tau) \dee\tau.
\end{align}
The sign difference between the two PID control laws arises from the dynamics in~\eqref{EOM_2}, where the AMT-induced torque $\mbs{\tau}_b^{\text{AMT}} = \f{m_s}{m_p+m_s}\mbf{r}_b^{ps^\times} \mbf{f}_b^{\text{SRP}}$ involves a cross product with opposite signs along the 1 and 2 axes.
The AMT control operates with a time step of $100$~sec. 
Between time steps, a ZOH is used instead of a FOH, since the PID controller lacks predictive capability for future inputs.

The on-off thresholds for AMT and RCDs, along with the PID gains for AMT control, are tuned based on simulation performance.
However, this control framework does not inherently account for physical actuator constraints, such as AMT position and rate limits, which are instead enforced after the PID controller determines the AMT position. 
Tuning the controller to ensure effective momentum management while avoiding actuator saturation remains a key challenge.

Unfortunately, the momentum management strategies outlined in~\cite{Inness2023,Tyler2024} are provided with very little detail on the numerical values used. Numerical values are chosen in this work in an attempt to recreate the results of~\cite{Inness2023,Tyler2024}. To this end, the chosen thresholds for the AMT are $0.125$~Nms for activation, and $0.0312$~Nms for deactivation.
The PID gains of the AMT controller are chosen as $K^\text{AMT}_p = 0.4$~(Ns)$^{-1}$, $K^\text{AMT}_d = 0.4$~N$^{-1}$, and $K^\text{AMT}_i = 0.0002$~N$^{-1}$s$^{-2}$.
The maximum position constraint of the AMT is enforced such that $|u^{\text{AMT}}_{i}| = u^{\text{AMT}}_{i,\text{max}} = 0.29$~m when the determined PID controller input satisfies $|u^{\text{AMT}}_{i}| > u^{\text{AMT}}_{i,\text{max}}$ ($i = 1, 2$).
The maximum AMT rate constraint is enforced such that $|u^{\text{AMT}}_{i}| = \Delta t \cdot \dot{u}^{\text{AMT}}_{i,\text{max}} = 0.05$~m when the determined PID input satisfies $|u^{\text{AMT}}_{i}| > \Delta t \cdot \dot{u}^{\text{AMT}}_{i,\text{max}}$ ($i = 1, 2$).
The chosen RCD thresholds are $0.25$~Nms for activation, and $0.125$~Nms for deactivation.

\begin{figure}[t!] 
\centering
\subfigure[momentum management inputs]
{
        \includegraphics[width=0.65\textwidth]{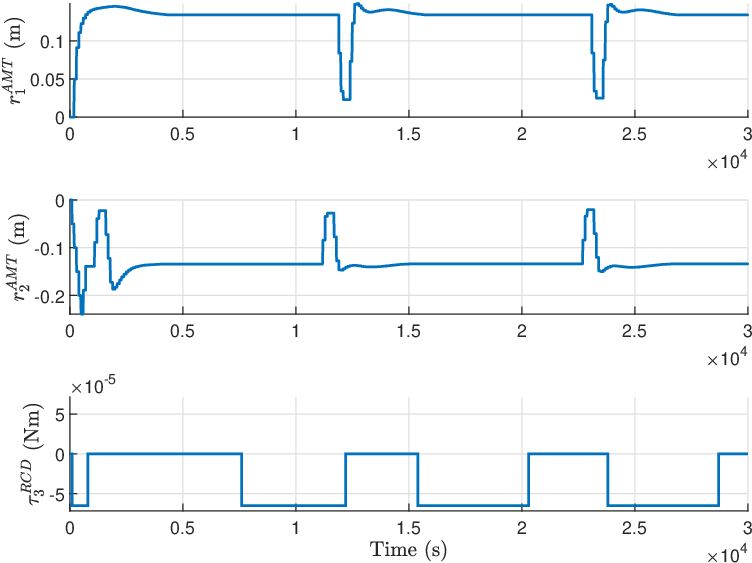}
        \label{Fig_NASA_sima}
}
\subfigure[RWs angular momentum]
{
        \includegraphics[width=0.65\textwidth]{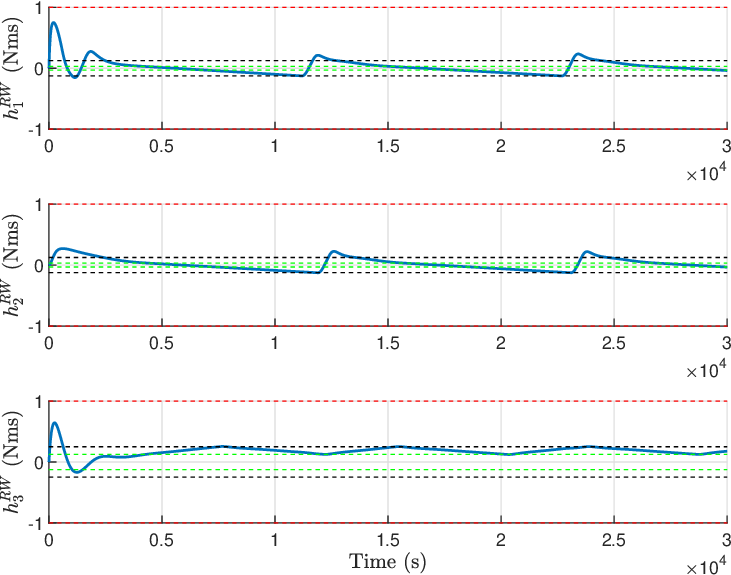}
        \label{Fig_NASA_simb}
}
\centering
\caption{Simulation results using NASA's Solar Cruiser momentum management strategy from~\cite{Inness2023,Tyler2024}.}
\label{Fig_NASA_sim}
\end{figure}

Simulation results with this momentum management strategy and the same initial conditions and parameters used in Section~\ref{sec:Sim_PID} are included in Figure~\ref{Fig_NASA_sim}. The momentum management performance is similar to that in~\cite{Inness2023,Tyler2024}, although this is difficult to compare quantitatively due to redacted plot axes in~\cite{Inness2023,Tyler2024}. The black dashed lines in Figure~\ref{Fig_NASA_simb} indicate the activation thresholds, while the green dashed lines represent the deactivation thresholds. The momentum management method from~\cite{Inness2023,Tyler2024} is effective at keeping the angular momentum of the RWs within reasonable bounds with realistic actuation inputs. However, it is worth noting that tuning the PID gains to obtain a satisfactory result required a significant amount of tuning.

\subsection{Proposed MPC-Based Momentum Management} \label{sec:Sim_MPC}


This section presents solar sail momentum management simulation results using the two MPC policies proposed in~\eqref{eq:MPC_cost2} and~\eqref{eq:MPC_cost_iter}. The same PID gains for the RWs and initial conditions as in Section~\ref{sec:Sim_PID} are used in these simulations.
The disturbance torque $\mbs{\tau}^{\text{dist}}_b$ is assumed to be perfectly known and provided to the MPC prediction model, resulting in $\mbf{w}_k = \mbs{\tau}^{\text{dist}}_b$ in the MPC formulation.
The lower bounds of the state constraints and input constraints are defined as the opposite of the upper bounds.
The state constraints in MPC are given by the spacecraft attitude drifting limit $\mbs{\theta}_{\text{max}} = -\mbs{\theta}_{\text{min}} = \Big[ 5\,\,\,5\,\,\,5\Big]^\trans$~deg, the angular rate limit $\mbs{\omega}^{ba}_{b,\text{max}} = -\mbs{\omega}^{ba}_{b,\text{min}} = \Big[ 20 \,\,\, 20 \,\,\, 20 \Big]^\trans$~deg/s, the RW angular momentum capacity $\mbf{h}^{\text{RWs}}_{b,\text{max}} = -\mbf{h}^{\text{RWs}}_{b,\text{min}} = \Big[ 1 \,\,\, 1  \,\,\,1\Big] ^\trans$~N$\cdot$m/s, and a large PID integral term $\mbf{e}^{\text{int}}_{\text{max}} = -\mbf{e}^{\text{int}}_{\text{min}} = \Big[ 10^6 \,\,\, 10^6  \,\,\, 10^6\Big] ^\trans$ as an internal state limit.
The input constraints in the MPC formulation are given by the actuator capabilities as outlined in Section~\ref{sec:Num_Setup} and Table~\ref{tab:table1}. The discrete rate constraint is defined as $\dot{\mbf{u}}^\text{AMT}_{\text{max}} = -\dot{\mbf{u}}^\text{AMT}_{\text{min}} = (\mbf{u}^\text{AMT}_{j+1|t_k} - \mbf{u}^\text{AMT}_{j|t_k})/\Delta t$, with $\Delta t = 100$~sec representing  the momentum management time step.
The MPC prediction horizon is chosen as $N = 20$ timesteps, corresponding to a $2000$~sec forecast.

The slack variable $\mbs{\alpha} \geq \mbf{0}$ is introduced to allow the RWs angular momentum to deviate from the chosen soft constraint bounds $\mbf{h}_{b,\text{min}}^\text{soft}$ and $\mbf{h}_{b,\text{max}}^\text{soft}$, where $\mbf{h}_{b,\text{min}}^\text{soft} - \mbs{\alpha} \leq \mbf{h}^\text{RWs}_{b,j|t_k} \leq \mbf{h}_{b,\text{max}}^\text{soft} + \mbs{\alpha}$.
This slack variable is penalized heavily by the weighting matrix $\mbf{C}$ in the cost function when $\mbf{h}^\text{RWs}_{b,j|t_k}$ deviates from the soft constraint region $\mbf{h}_{b,\text{min}}^\text{soft} \leq \mbf{h}^\text{RWs}_{b,j|t_k} \leq \mbf{h}_{b,\text{max}}^\text{soft}$.
The values of soft constraints are chosen as $25\%$ of the RWs angular momentum capacity, \ie, $\mbf{h}_{b,\text{max}}^\text{soft} = 0.25\cdot \mbf{h}^{\text{RWs}}_{b,\text{max}}$ and $\mbf{h}_{b,\text{min}}^\text{soft} = -\mbf{h}_{b,\text{max}}^\text{soft}$.
The MPC tuning parameters are chosen as $\mbf{Q}=\text{diag}(10\cdot\mbf{1}_{6\times6}, 10^{-2}, 10^{-2}, 10^{-8}, \mbf{0}_{3\times3})$, $\mbf{R} =\text{diag}(1,1,10^6)$, $\tilde{\mbf{R}} = \text{diag}(10, 10)$, $\mbf{C} = 10^3\cdot\mbf{1}_{3\times3}$, and the terminal costs $\mbf{P} = \mbf{1}_{12\times12}$, $\mbf{R}_N = \mbf{R}$.
In the simulation, the linear interpolation of the FOH input ($\mbf{u}^\text{AMT}$) is computed using equation~\eqref{eq:input_DTLTI} with the RW time step of $\dee t = 1$~sec, while $\tau^\text{RCD}_{b3}$ and $\mbs{\tau}^\text{dist}_b$ are held as constant between momentum management timesteps according to ZOH.

\begin{figure}[t!] 
\centering
\subfigure[momentum management inputs]
{
        \includegraphics[width=0.65\textwidth]{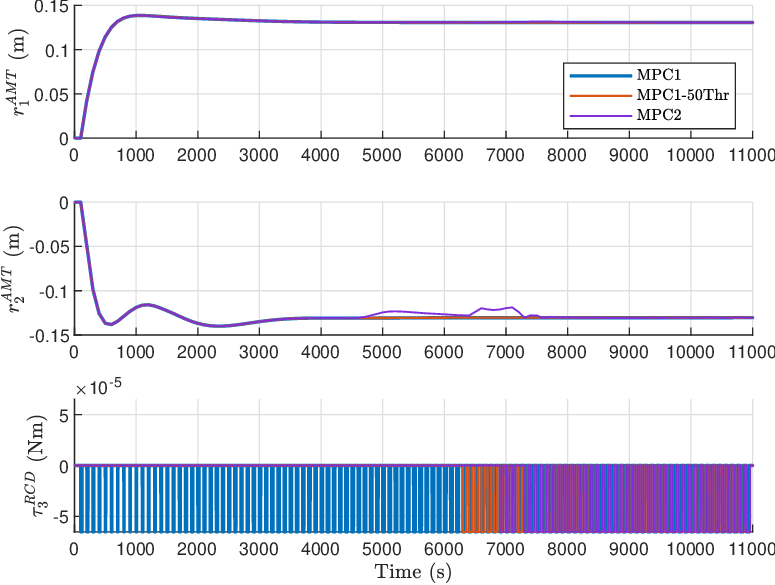}
        \label{Fig_full_sima}
}
\subfigure[RWs angular momentum]
{
        \includegraphics[width=0.65\textwidth]{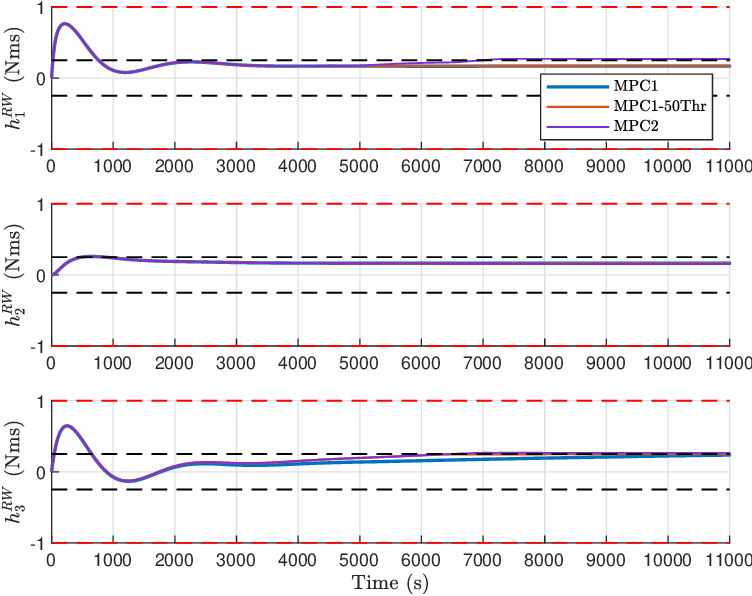}
        \label{Fig_full_simb}
}
\centering
\caption{Complete simulation results with $50\%$ thresholds on RCD inputs with non-iterative MPC (MPC Strategy~1) as well as backwards-iterative method with RCD threshold (MPC Strategy~2). }
\label{Fig_full_sim}
\end{figure}

Figure~\ref{Fig_full_sim} presents a comparison of the simulation results of three MPC momentum management controllers using the same tuning. 
The first and second MPC controllers are based on the LQ-MPC policy in~\eqref{eq:MPC_cost2} (MPC Strategy~1) with and without the additional RCD threshold.
The threshold serves as another tuning knob that reduces actuator usage until a sufficiently large input is required. When the MPC controller calculates an RCD input below the threshold, it is ignored; larger inputs are quantized using a PWM-based method as in Section~\ref{sec:PWM} before being applied.
The first controller, using PWM-quantized RCD inputs without any threshold, is shown by the blue line, while the second controller, which applies a $50\%$ threshold of the maximum RCD torque value, is represented by the red line. 
In other words, the second controller with RCD threshold forces $\tau^\text{RCD}_{b3} = 0$ when $\tau^\text{RCD}_{b3} < 0.5\cdot u^\text{RCD}_\text{max}$. 
The third controller, using the iterative MPC policy from equation~\eqref{eq:MPC_cost_iter} (MPC Strategy~2), is indicated by the purple line, with both the RCD threshold and PWM-quantization embedded into the fixed virtual input throughout the iterations. 
Figure~\ref{Fig_full_sima} shows the momentum management actuator inputs, where an ideal performance is to reduce the usage of momentum management actuators with minimal number of RCD on-off cycles, RCD ``on'' time, and AMT moving distance, while successfully managing RWs angular momentum.
Figure~\ref{Fig_full_simb} shows the RW angular momentum in all axes, where the black dashed lines are the soft constraints, and the red dashed lines are the hard constraints based on the RW saturation limits. 
There is a significant discrepancy in RCD usage as well as a slight RWs angular momentum history difference between the three MPC momentum management controllers.

\begin{figure}[t!] 
\centering
\subfigure[momentum management inputs]
{
        \includegraphics[width=0.65\textwidth]{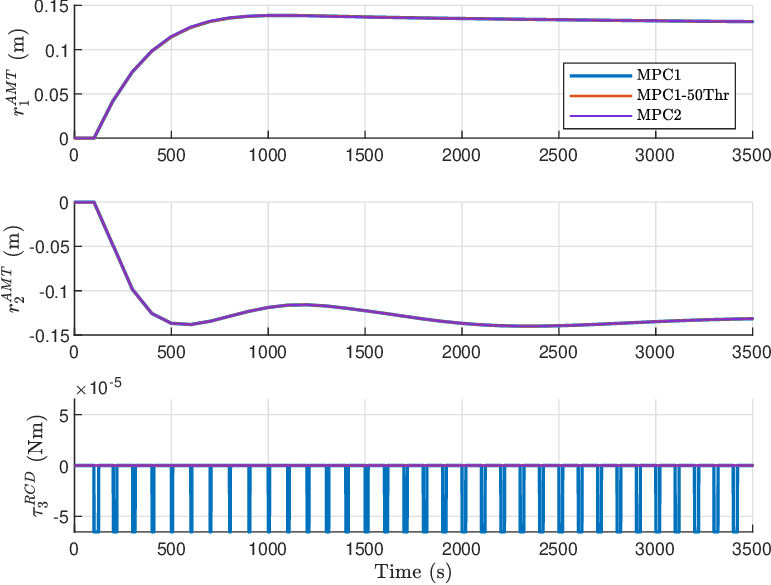}
}
\subfigure[RWs angular momentum]
{
        \includegraphics[width=0.65\textwidth]{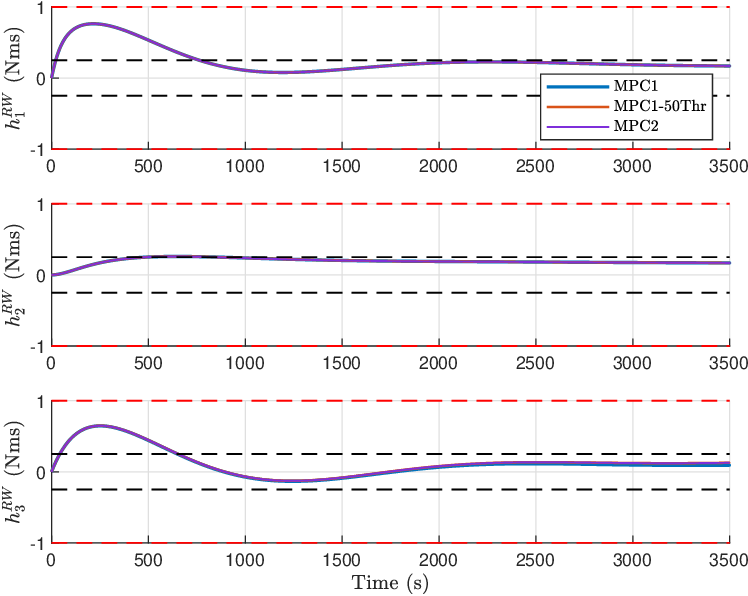}
}
\centering
\caption{Initial maneuver state comparison of adding $50\%$ thresholds on RCD inputs with non-iterative MPC (MPC Strategy~1) as well as backwards-iterative method with RCD threshold (MPC Strategy~2).}
\label{Fig_initial_sim}
\end{figure}

\begin{figure}[t!] 
\centering
\subfigure[momentum management inputs]
{
        \includegraphics[width=0.65\textwidth]{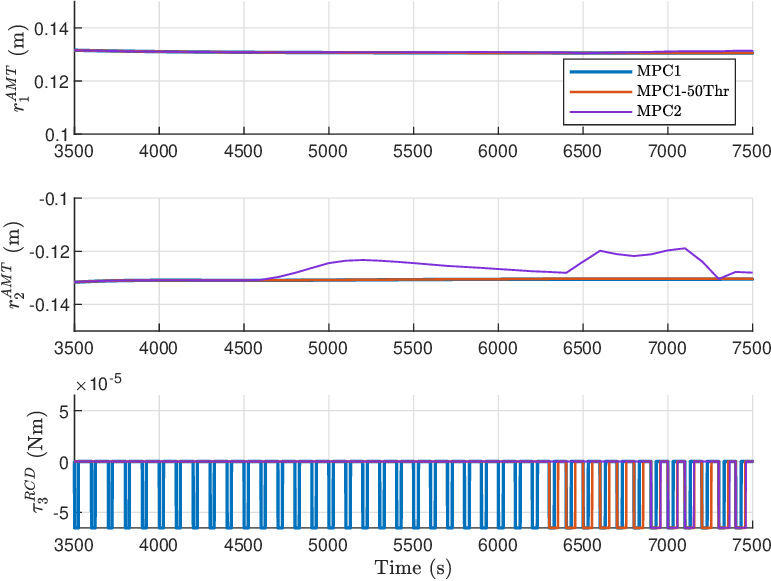}
}
\subfigure[RWs angular momentum]
{
        \includegraphics[width=0.65\textwidth]{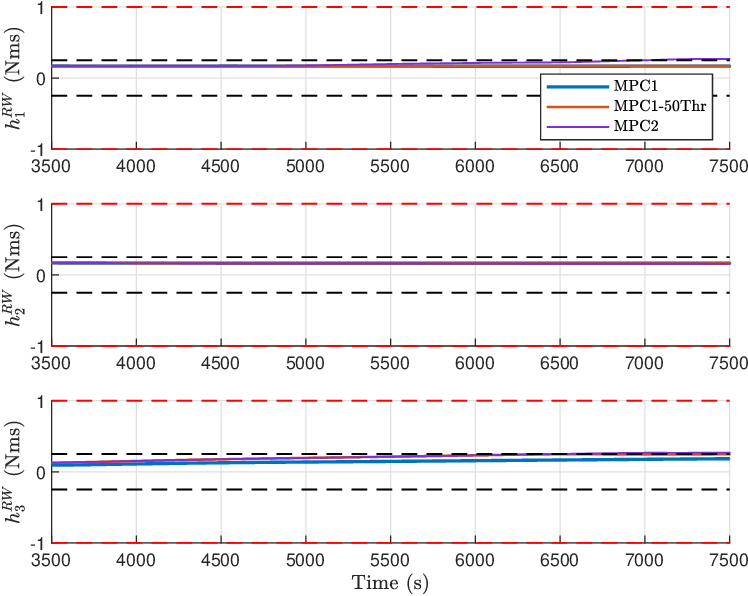}
}
\centering
\caption{Transient stage comparison of adding $50\%$ thresholds on RCD inputs with non-iterative MPC (MPC Strategy~1) as well as backwards-iterative method with RCD threshold (MPC Strategy~2).}%
\label{Fig_transient_sim}
\end{figure}

\begin{figure}[t!] 
\centering
\subfigure[momentum management inputs]
{
        \includegraphics[width=0.65\textwidth]{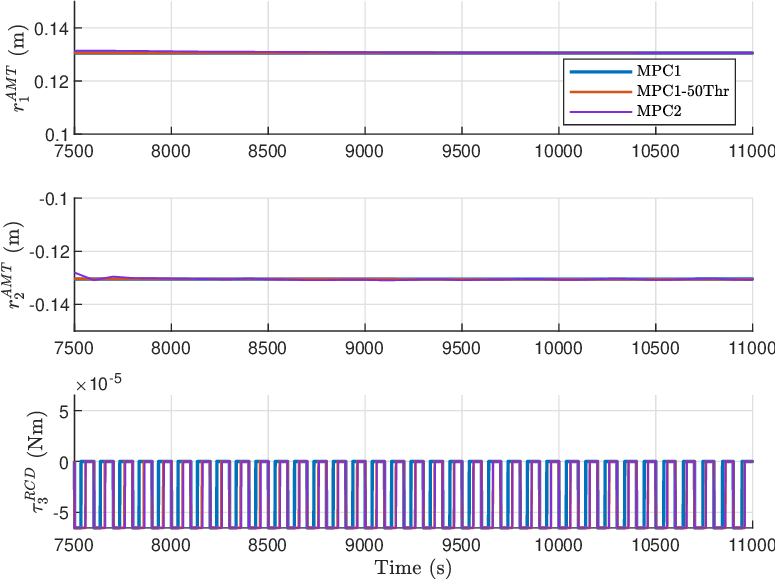}  
}
\subfigure[RWs angular momentum]
{
        \includegraphics[width=0.65\textwidth]{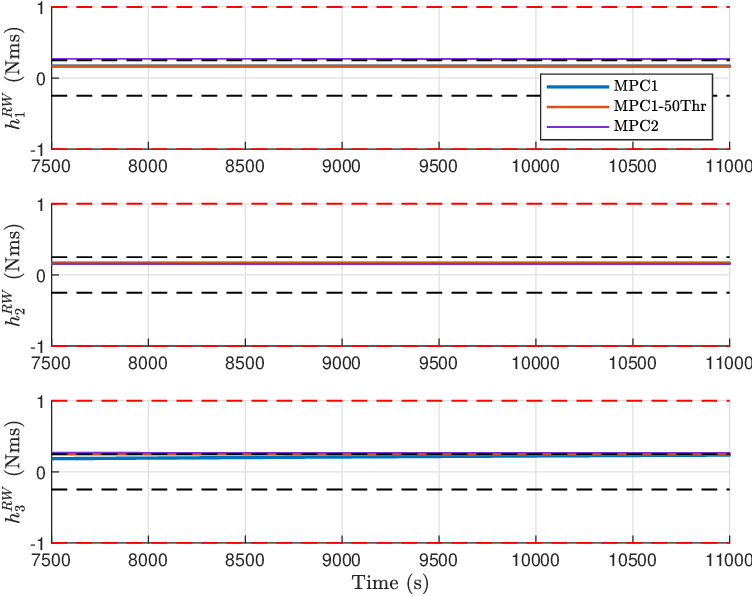}
}
\centering
\caption{Steady state comparison of adding $50\%$ thresholds on RCD inputs with non-iterative MPC (MPC Strategy~1) as well as backwards-iterative method with RCD threshold (MPC Strategy~2).
}\label{Fig_steady_sim}
\end{figure}

To better analyze the results, the simulation is dissected into 3 stages of operational condition (approximately one hour each), where the initial maneuver between time $0-3500$~sec, the transient state between time $3500-7500$~sec, and the steady state between time $7500-11000$~sec.
The momentum management input and RWs angular momentum history of each stage are shown in Figs~\ref{Fig_initial_sim},~\ref{Fig_transient_sim}, and~\ref{Fig_steady_sim}.
Quantitative data for each MPC controller's actuation usage in the respective operational stages are provided in Tables~\ref{tab:sim_data_initial_maneuver},~\ref{tab:sim_data_transient}, and~\ref{tab:sim_data_steady_state}.
AMT and RCDs actuation usage of Solar Cruiser's state-of-the-art method are also presented as a comparison.


In the initial stage ($0 - 3500$~sec), as shown in Figure~\ref{Fig_initial_sim} and Table~\ref{tab:sim_data_initial_maneuver}, the RWs quickly accumulate angular momentum as they conduct the maneuver to track the desired attitude. 
At $t = 100$~sec, the MPC controllers are activated, and all controllers successfully manage the RWs angular momentum while maintaining attitude tracking. The first MPC controller without an RCD threshold (blue line) utilizes $34$ RCD on-off cycles, while the second and third controllers (with RCD threshold and backwards-iterative MPC, respectively) reduce RCD usage significantly, requiring $0$ cycles.
Compared to Solar Cruiser's PID-based momentum management strategy, the actuator usage in RCD time and AMT travel distance are significantly lower during the initial trajectory tracking stage.

During the transient stage ($3500 - 7500$~sec), as shown in Figure~\ref{Fig_transient_sim} and Table~\ref{tab:sim_data_transient}, the AMT is moving towards its optimal position to counteract the constant disturbance, and the roll axis RW angular momentum stabilizes with RCD actuation.
The backwards-iterative MPC controller (purple line) extends the transient stage and delays the onset of steady-state RCD actuation, which results in fewer RCD on-off cycles, but with a larger AMT travel distance.
Specifically, the backwards-iterative MPC controller exhibits small AMT movements when entering the steady-state operation period. This can be due to the necessity of continuous RCD actuation in steady state to counteract the constant disturbance, which results in an imbalance of the input weights in the MPC cost function.
Considering the trade-off between AMT usage and RCD actuation, fewer RCD cycles and RCD time are more preferable, as the RCD is a relatively new technology with uncertain lifespan due to issues such as overheating and being rated for a limited number of on-off cycles.

In the steady-state results of Figure~\ref{Fig_steady_sim} and Table~\ref{tab:sim_data_steady_state}, all three MPC controllers converge to the same steady-state RCD cycle, and the AMT position stabilizes. 
The backwards-iterative MPC controller, however, continues to exhibit small AMT movements, which is not ideal. 
In this case, the second controller (non-iterative MPC with RCD threshold) provides the best performance, maintaining minimal AMT motion and RCD usage.


\begin{table}[t!]
    \caption{Initial maneuver momentum management actuator usage for time $0 - 3500$~sec.}
    \centering
    \begin{tabular}{ccccc}
         & RCD On-Off& RCD & AMT Distance & AMT Distance \\
         & Cycles  & On Time & Traveled in $\vect{b}^1$ & Traveled in $\vect{b}^2$\\
         & (times) & (sec) & (cm) & (cm)\\
        MPC1& 34 & 515 & 14.5574 & 19.2945\\
        MPC1-Thr& 0 & 0 & 14.5528 & 19.2927\\
        MPC2& 0 & 0 & 14.5390 & 19.2883\\
        Solar Cruiser & 1 & 700 & 15.2928  & 67.0687\\
    \end{tabular}
    \label{tab:sim_data_initial_maneuver}
\end{table}


\begin{table}[t!]
    \caption{Transient state momentum management actuator usage for time $3500 - 7500$~sec.}
    \centering
    \begin{tabular}{ccccc}
         & RCD On-Off& RCD & AMT Distance & AMT Distance \\
         & Cycles  & On Time & Traveled in $\vect{b}^1$ & Traveled in $\vect{b}^2$\\
         & (times) & (sec) & (cm) & (cm)\\
        MPC1& 40 & 1098 & 0.1043 & 0.1222\\
        MPC1-Thr& 12 & 653 & 0.1040 & 0.1444\\
        MPC2& 5 & 291 & 0.2445 & 4.0863\\
        Solar Cruiser & 0 & 0 & 0.3418  & 0.2006\\
    \end{tabular}
    \label{tab:sim_data_transient}
\end{table}


\begin{table}[t!]
    \caption{Steady-state momentum management actuator usage for time $7500 - 11000$~sec.}
    \centering
    \begin{tabular}{ccccc}
         & RCD On-Off& RCD & AMT Distance & AMT Distance \\
         & Cycles  & On Time & Traveled in $\vect{b}^1$ & Traveled in $\vect{b}^2$\\
         & (times) & (sec) & (cm) & (cm)\\
        MPC1& 35 & 1286 & 0.0009 & 0.0078\\
        MPC1-Thr& 35 & 2031 & 0.0012 & 0.0147\\
        MPC2 & 35 & 2094 & 0.0888  & 0.9647\\
        Solar Cruiser & 0.5 & 3400 & 0  & 0\\
    \end{tabular}
    \label{tab:sim_data_steady_state}
\end{table}

This numerical simulation demonstrates the strengths and benefits of the proposed MPC momentum management strategies under various operational conditions. The backwards-iterative MPC strategy performs well during the initial maneuver and transient stages, with a significant reduction in RCD usage, although it requires more AMT movement and computational power. The PWM-quantized MPC with the RCD threshold provides the best steady-state performance and is also effective during the initial maneuver and transient stages when computational resources are limited.


\subsection{Discussion and Robustness Assessment}

A comparison of $30000$ seconds (approximately $8$ hours) simulation between the Solar Cruiser PID-based momentum management method and the proposed MPC-based approach is provided in Table~\ref{tab:sim_data_NASA_comp}. 
Compared to the state-of-the-art momentum management strategy used on Solar Cruiser, the proposed MPC-based controller in Section~\ref{sec:MPC} effectively manages momentum while significantly reducing actuator usage. 
The substantial reduction in total AMT travel distance demonstrates improved efficiency, which conserves power resources and enhances actuator longevity.
Although the MPC-based approaches result in a larger number of RCD on-off cycles compared to Solar Cruiser’s method, the total RCD on time is lower.

A potential concern with the proposed MPC strategy is the excitation of the solar sail's structural frequency modes due to its on-off actuation. To alleviate this concern, an analysis is conducted comparing the system's natural frequencies to the actuation frequency. Based on the dynamic model developed in~\cite{Bunker2024} with publicly-available parameters of the TRAC booms in~\cite{nguyen2023solar}, it is estimated that the lowest natural frequency of the system is $0.1150$~Hz. 
Given the actuation frequency of $0.01$~Hz of the proposed MPC-based methods being an order of magnitude lower than the lowest frequency mode, the slow actuation rate is unlikely to excite the flexible modes of the structure.

\begin{table}[t!]
    \caption{Comparison of actuator usage between NASA Solar Cruiser's momentum management strategy and the proposed MPC-based synthesis from $0$ to $30000$ sec.}
    \centering
    \begin{tabular}{ccccc}
         & RCD On-Off& RCD & AMT Distance & AMT Distance \\
         & Cycles  & On Time & Traveled in $\vect{b}^1$ & Traveled in $\vect{b}^2$\\
         & (times) & (sec) & (cm) & (cm)\\
        MPC1 & 299 & 13884 & 14.6607 & 19.4408\\
        MPC1-Thr & 237 & 13884 & 14.6582  & 19.4551\\
        MPC2 & 230 & 13589 & 14.9879  & 26.5160\\
        Solar Cruiser & 4 & 15100 & 66.6281  & 118.3571\\
    \end{tabular}
    \label{tab:sim_data_NASA_comp}
\end{table}

\begin{figure}[t!] 
        \centering
        \subfigure[momentum management inputs]
        {
                \includegraphics[width=0.65\textwidth]{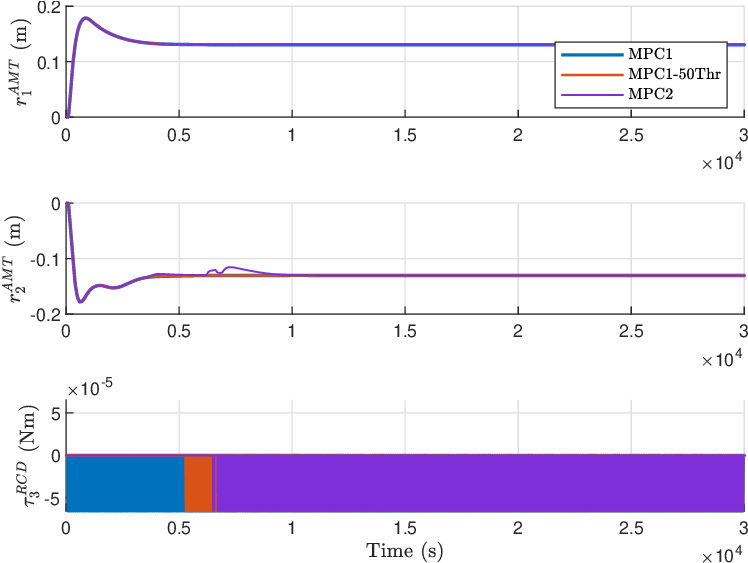}
        }
        \subfigure[RWs angular momentum]
        {
                \includegraphics[width=0.65\textwidth]{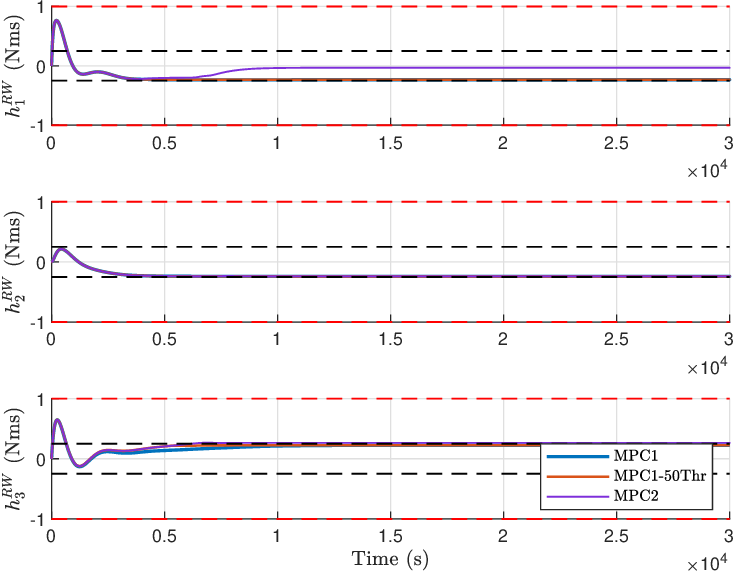}
        }
        \centering
        \caption{The proposed MPC strategies featuring $50\%$ overestimate of disturbance, where $\mbf{w}_k = 1.5 \mbs{\tau}^\text{dist}_b$.}
        \label{Fig_30overEst}
        \end{figure}

\begin{figure}[t!] 
        \centering
        \subfigure[momentum management inputs]
        {
                \includegraphics[width=0.65\textwidth]{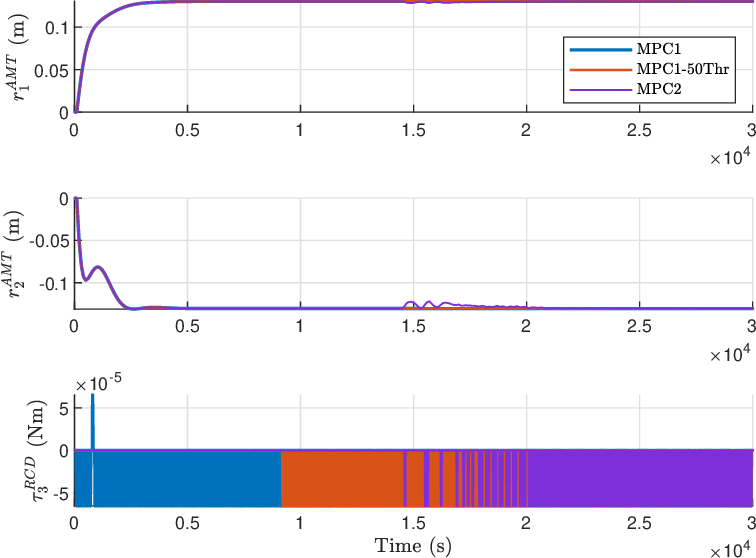}
        }
        \subfigure[RWs angular momentum]
        {
                \includegraphics[width=0.65\textwidth]{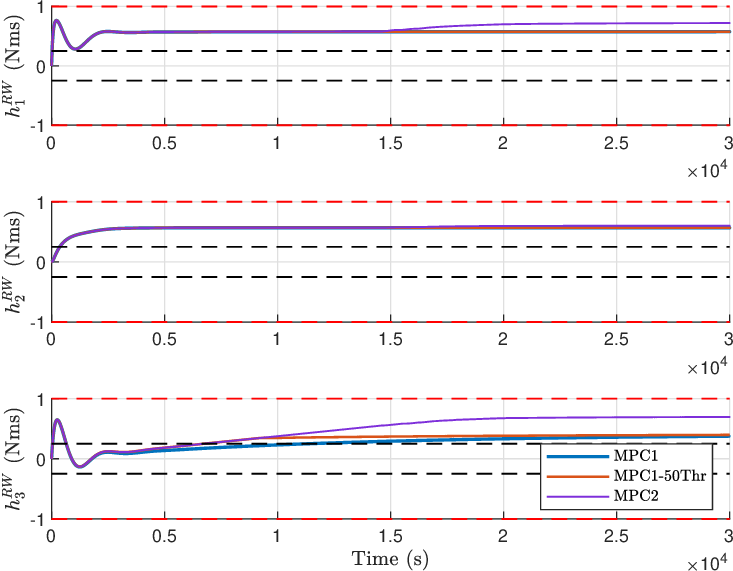}
        }
        \centering
        \caption{The proposed MPC strategies featuring $50\%$ underestimate of disturbance, where $\mbf{w}_k = 0.5 \mbs{\tau}^\text{dist}_b$.}
        \label{Fig_10underEst}
        \end{figure}

To further assess robustness of the proposed MPC-based controllers, uncertainty of $-50\%$ and $+50\%$ in the expected disturbance $\mbf{w}_k$ used within the MPC prediction model is evaluated, as shown in Figures~\ref{Fig_30overEst} and~\ref{Fig_10underEst}. This range of disturbance error is a conservative representative of the amount of error to be expected based on measurements available onboard the solar sail, as shown in~\cite{diedrich2023solar}. 
Under these conditions, the performance remain stable without constraint violation. 
The MPC Strategy 2 iteratively optimizes the control input based on the knowledge of prediction model, resulting in an amplification of the underestimated disturbance error.
Practically, it appears to be more preferable to overestimate the disturbance in the MPC than underestimating it.

\section{Conclusion}

The novel MPC-based momentum management strategies presented in this paper demonstrated the ability to handle the coupled nature of the solar sail's dynamics in a practical manner while minimizing actuator usage and respecting the limited onboard computation resources. 
Simulation results showed the effectiveness of the proposed MPC-based momentum management strategies in the presence of a constant worst-case disturbance torque.
The proposed MPC strategies feature significant improvement in reducing momentum management actuation usage in contrast to the state-of-the-art decoupled controller framework while preserving simplicity for onboard implementation.
The iterative backwards-in-time MPC algorithm further captures the RCD's on-off actuation and dead-band thresholds in the prediction model, and improved actuation efficiency when computation capacity is allowed.

Future work will focus on improving the MPC prediction model accuracy by incorporating state propagation or iterative linear time-varying (LTV) dynamics. Additionally, a disturbance estimation framework will be developed to address the lack of knowledge of the disturbance in practice.


\section*{Acknowledgments}

This work was supported in part by a study grant from Chung Cheng Institute of Technology, National Defense University, Taiwan (R.O.C.), as well as funding from the Research \& Innovation Office, University of Minnesota.

\bibliography{sample_nodoi}

\begin{thebibliography}{10}
\expandafter\ifx\csname url\endcsname\relax
  \def\url#1{\texttt{#1}}\fi
\expandafter\ifx\csname urlprefix\endcsname\relax\def\urlprefix{URL }\fi
\expandafter\ifx\csname href\endcsname\relax
  \def\href#1#2{#2} \def\path#1{#1}\fi

\bibitem{berthet2024space}
M.~Berthet, J.~Schalkwyk, O.~{\c C}elik, D.~Sengupta, K.~Fujino, A.~M. Hein,
  L.~Tenorio, J.~Cardoso~dos Santos, S.~P. Worden, P.~D. Mauskopf, Y.~Miyazaki,
  I.~Funaki, S.~Tsuji, P.~Fil, K.~Suzuki, Space sails for achieving major space
  exploration goals: Historical review and future outlook, Progress in
  Aerospace Sciences 150 (2024) 101047.

\bibitem{zhang2023long}
J.~Zhang, Y.~Yuan, K.~Yang, L.~Li, Long-term evolution of the space environment
  considering constellation launches and debris disposal, IEEE Transactions on
  Aerospace and Electronic Systems 59~(5) (2023) 6124--6137.

\bibitem{bianchi2024preliminary}
C.~Bianchi, L.~Niccolai, G.~Mengali, M.~Ceriotti, Preliminary design of a space
  debris removal mission in leo using a solar sail, Advances in Space Research
  73~(8) (2024) 4254--4268.

\bibitem{bigdeli2025mechanics}
M.~Bigdeli, R.~Srivastava, M.~Scaraggi, Mechanics of space debris removal: A
  review, Aerospace 12~(4) (2025) 277.

\bibitem{guglielmo2019drag}
D.~Guglielmo, S.~Omar, R.~Bevilacqua, L.~Fineberg, J.~Treptow, B.~Poffenberger,
  Y.~Johnson, Drag deorbit device: A new standard reentry actuator for
  cubesats, Journal of Spacecraft and Rockets 56~(1) (2019) 129--145.

\bibitem{nagavarapu2025cubesats}
S.~C. Nagavarapu, L.~B. Mogan, A.~Chandran, D.~E. Hastings, Cubesats for space
  debris removal from leo: Prototype design of a robotic arm-based deorbiter
  cubesat, Advances in Space Research 75~(9) (2025) 6896--6910.

\bibitem{eldad2015propellantless}
O.~Eldad, E.~G. Lightsey, Propellantless attitude control of a nonplanar solar
  sail, Journal of Guidance, Control, and Dynamics 38~(8) (2015) 1531--1534.

\bibitem{farres2023propellant}
A.~Farres, Propellant-less systems, in: F.~Branz, C.~Cappelletti, A.~J. Ricco,
  J.~W. Hines (Eds.), Next Generation CubeSats and SmallSats, Elsevier, 2023,
  pp. 519--541.

\bibitem{miller2022high}
D.~Miller, F.~Duvigneaud, W.~Menken, D.~Landau, R.~Linares, High-performance
  solar sails for interstellar object rendezvous, Acta Astronautica 200 (2022)
  242--252.

\bibitem{JohnsonLes2020SCTM}
L.~Johnson, F.~Curran, {Solar Cruiser} technology maturation plans, Tech. Rep.
  20205003681, NASA Marshall Space Flight Center (2020).

\bibitem{johnson2019solar}
L.~Johnson, F.~M. Curran, R.~Dissly, A.~F. Heaton, The {Solar Cruiser} mission:
  Demonstrating large solar sails for deep space missions, in: International
  Astronautical Congress, no. MSFC-E-DAA-TN74364, Washington, DC, 2019.

\bibitem{pezent2021preliminary}
J.~B. Pezent, R.~Sood, A.~Heaton, K.~Miller, L.~Johnson, Preliminary trajectory
  design for {NASA's Solar Cruiser}: A technology demonstration mission, Acta
  Astronautica 183 (2021) 134--140.

\bibitem{johnson2022nasa}
L.~Johnson, J.~Everett, D.~McKenzie, D.~Tyler, D.~Wallace, J.~Newmark,
  D.~Turse, M.~Cannella, M.~Feldman, The nasa solar cruiser mission - solar
  sail propulsion enabling heliophysics missions, in: 36th Annual Small
  Satellite Conference, 2022.

\bibitem{gauvain2023solar}
B.~M. Gauvain, D.~A. Tyler, A solar sail shape modeling approach for attitude
  control design and analysis, in: 6th International Symposium on Space
  Sailing, New York, NY, 2023.

\bibitem{fu2015attitude}
B.~Fu, F.~O. Eke, Attitude control methodology for large solar sails, Journal
  of Guidance, Control, and Dynamics 38~(4) (2015) 662--670.

\bibitem{firuzi2018attitude}
S.~Firuzi, S.~Gong, Attitude control of a flexible solar sail in low {Earth}
  orbit, Journal of Guidance, Control, and Dynamics 41~(8) (2018) 1715--1730.

\bibitem{wie2004solar1}
B.~Wie, Solar sail attitude control and dynamics, part 1, Journal of Guidance,
  Control, and Dynamics 27~(4) (2004) 526--535.

\bibitem{wie2004solar2}
B.~Wie, Solar sail attitude control and dynamics, part 2, Journal of Guidance,
  Control, and Dynamics 27~(4) (2004) 536--544.

\bibitem{ingrassia2013solar}
T.~Ingrassia, V.~Faccin, A.~Bolle, C.~Circi, S.~Sgubini, Solar sail elastic
  displacement effects on interplanetary trajectories, Acta Astronautica 82~(2)
  (2013) 263--272.

\bibitem{Inness2023}
J.~Inness, D.~Tyler, B.~Diedrich, S.~Ramazani, J.~Orphee, Momentum management
  strategies for {Solar Cruiser} and beyond, in: 6th International Symposium on
  Space Sailing, New York, NY, 2023.

\bibitem{inness2024controls}
J.~Inness, B.~Diedrich, B.~Valdez, D.~Tyler, B.~Sanders, Controls modeling
  approach for deployment of a large thin structures for solar sails, in: 38th
  Annual Small Satellite Conference, Logan, UT, 2024.

\bibitem{guerrant2015tactics}
D.~Guerrant, D.~Lawrence, Tactics for heliogyro solar sail attitude control via
  blade pitching, Journal of Guidance, Control, and Dynamics 38~(9) (2015)
  1785--1799.

\bibitem{choi2016control}
M.~Choi, C.~J. Damaren, Control allocation of solar sail tip vanes with two
  degrees of freedom, Journal of Guidance, Control, and Dynamics 39~(8) (2016)
  1857--1865.

\bibitem{abrishami2020optimized}
A.~Abrishami, S.~Gong, Optimized control allocation of an articulated
  overactuated solar sail, Journal of Guidance, Control, and Dynamics 43~(12)
  (2020) 2321--2332.

\bibitem{hassanpour2020collocated}
S.~Hassanpour, C.~J. Damaren, Collocated attitude and vibrations control for
  square solar sails with tip vanes, Acta Astronautica 166 (2020) 482--492.

\bibitem{Tyler2024}
D.~Tyler, B.~Diedrich, B.~Gauvain, J.~Inness, A.~Heaton, J.~Orphee, Attitude
  control approach for {Solar Cruiser}, a large, deep space solar sail mission,
  in: AAS Guidance, Navigation and Control Conference, Breckenridge, CO, 2023.

\bibitem{camacho2007constrained}
E.~F. Camacho, C.~Bordons, Constrained Model Predictive Control, Springer,
  London, UK, 2007, pp. 177--216.

\bibitem{rawlings2017model}
J.~B. Rawlings, D.~Q. Mayne, M.~Diehl, Model Predictive Control: Theory,
  Computation, and Design, Vol.~2, Nob Hill Publishing, Madison, WI, 2017.

\bibitem{morari1999model}
M.~Morari, J.~H. Lee, Model predictive control: Past, present and future,
  Computers \& Chemical Engineering 23~(4-5) (1999) 667--682.

\bibitem{mayne2014model}
D.~Q. Mayne, Model predictive control: Recent developments and future promise,
  Automatica 50~(12) (2014) 2967--2986.

\bibitem{eren2017model}
U.~Eren, A.~Prach, B.~B. Ko{\c{c}}er, S.~V. Rakovi{\'c}, E.~Kayacan,
  B.~A{\c{c}}{\i}kme{\c{s}}e, Model predictive control in aerospace systems:
  Current state and opportunities, Journal of Guidance, Control, and Dynamics
  40~(7) (2017) 1541--1566.

\bibitem{di2018real}
S.~Di~Cairano, I.~V. Kolmanovsky, Real-time optimization and model predictive
  control for aerospace and automotive applications, in: American Control
  Conference, Milwaukee, WI, 2018, pp. 2392--2409.

\bibitem{petersen2023safe}
C.~Petersen, R.~J. Caverly, S.~Phillips, A.~Weiss, Safe and constrained
  rendezvous, proximity operations, and docking, in: American Control
  Conference, San Diego, CA, 2023, pp. 3645--3661.

\bibitem{lee2017geometric}
D.~Y. Lee, R.~Gupta, U.~V. Kalabi{\'c}, S.~Di~Cairano, A.~M. Bloch, J.~W.
  Cutler, I.~V. Kolmanovsky, Geometric mechanics based nonlinear model
  predictive spacecraft attitude control with reaction wheels, Journal of
  Guidance, Control, and Dynamics 40~(2) (2017) 309--319.

\bibitem{MAMMARELLA2018585}
M.~Mammarella, E.~Capello, H.~Park, G.~Guglieri, M.~Romano, Tube-based robust
  model predictive control for spacecraft proximity operations in the presence
  of persistent disturbance, Aerospace Science and Technology 77 (2018)
  585--594.

\bibitem{caverly2020electric}
R.~J. Caverly, S.~Di~Cairano, A.~Weiss, Electric satellite station keeping,
  attitude control, and momentum management by {MPC}, IEEE Transactions on
  Control Systems Technology 29~(4) (2020) 1475--1489.

\bibitem{jin2022model}
L.~Jin, Y.~Li, Model predictive control-based attitude control of
  under-actuated spacecraft using solar radiation pressure, Aerospace 9~(9)
  (2022) 498.

\bibitem{Halverson2024}
R.~D. Halverson, D.~Gebre-Egziabher, R.~J. Caverly, Attitude control of
  dual-spin satellites in low-{Earth} orbit via predictive control and magnetic
  actuation, in: AIAA SciTech Forum, Orlando, FL, 2024, {AIAA} 2024-2278.

\bibitem{hayes2023model}
A.~D. Hayes, R.~J. Caverly, Model predictive tracking of spacecraft deorbit
  trajectories using drag modulation, Acta Astronautica 202 (2023) 670--685.

\bibitem{martins2021engineering}
J.~R. Martins, A.~Ning, Engineering Design Optimization, Cambridge University
  Press, 2021.

\bibitem{axehill2006mixed}
D.~Axehill, A.~Hansson, A mixed integer dual quadratic programming algorithm
  tailored for {MPC}, in: IEEE Conference on Decision and Control, San Diego,
  CA, 2006, pp. 5693--5698.

\bibitem{axehill2010improved}
D.~Axehill, M.~Morari, Improved complexity analysis of branch and bound for
  hybrid {MPC}, in: IEEE Conference on Decision and Control, Atlanta, GA, 2010,
  pp. 4216--4222.

\bibitem{botelho2024explicit}
A.~Botelho, P.~Rosa, J.~M. Lemos, Explicit spacecraft thruster control
  allocation with minimum impulse bit, IEEE Transactions on Control Systems
  TechnologyIn press (2024).

\bibitem{chegeni2014attitude}
E.~Chegeni, M.~Zandieh, J.~Ebrahimi, Attitude control of satellite with
  pulse-width pulse-frequency ({PWPF}) modulator using generalized incremental
  predictive control, Majlesi Journal of Electrical Engineering 8~(3) (2014)
  25--31.

\bibitem{zlotnik2017mpc}
D.~Zlotnik, S.~Di~Cairano, A.~Weiss, {MPC} for coupled station keeping,
  attitude control, and momentum management of geo satellites using on-off
  electric propulsion, in: IEEE Conference on Control Technology and
  Applications, Maui, HI, 2017, pp. 1835--1840.

\bibitem{caverly2018off}
R.~J. Caverly, S.~Di~Cairano, A.~Weiss, On-off quantization of an {MPC} policy
  for coupled station keeping, attitude control, and momentum management of
  {GEO} satellites, in: European Control Conference, Limassol, Cyprus, 2018,
  pp. 3114--3119.

\bibitem{schaub2018analytical}
H.~Schaub, J.~L. Junkins, Analytical Mechanics of Space Systems, 4th Edition,
  AIAA, Reston, VA, 2018.

\bibitem{DeRuiter2013}
A.~H.~J. de~Ruiter, C.~J. Damaren, J.~R. Forbes, Spacecraft Dynamics and
  Control: An Introduction, John Wiley and Sons, Chichester, UK, 2013.

\bibitem{hayes2022atmospheric}
A.~D. Hayes, R.~J. Caverly, D.~Gebre-Egziabher, Atmospheric density estimation
  in low-{Earth} orbit for drag-modulated spacecraft, in: M.~Sandnas, D.~B.
  Spencer (Eds.), Proceedings of the 44th Annual American Astronautical Society
  Guidance, Navigation, and Control Conference, 2022, Springer, Cham,
  Switzerland, 2024, pp. 1179--1194.

\bibitem{de1994constraint}
N.~M. De~Oliveira, L.~T. Biegler, Constraint handing and stability properties
  of model-predictive control, AIChE Journal 40~(7) (1994) 1138--1155.

\bibitem{kerrigan2000soft}
E.~C. Kerrigan, J.~M. Maciejowski, Soft constraints and exact penalty functions
  in model predictive control, in: Control 2000 Conference, Cambridge, UK,
  2000, pp. 2319--2327.

\bibitem{heaton2023reflectivity}
A.~Heaton, S.~Ramazani, D.~Tyler, Reflectivity control device ({RCD}) roll
  momentum management for solar cruiser and beyond, in: 6th International
  Symposium on Solar Sailing, New York, NY, 2023.

\bibitem{BCTRW}
Blue canyon technologies reaction wheels data sheet,
  \url{https://storage.googleapis.com/blue-canyon-tech-news/1/2023/04/ReactionWheels.pdf},
  accessed: 2023-12-18 (2023).

\bibitem{heaton2015update}
A.~Heaton, A.~Artusio-Glimpse, An update to the {NASA} reference solar sail
  thrust model, in: AIAA SPACE Conference and Exposition, Pasadena, CA, 2015,
  {AIAA} 2015-4506.

\bibitem{Bunker2024}
K.~R. Bunker, R.~J. Caverly, Modular dynamic modeling and simulation of a
  cable-actuated flexible solar sail, in: AIAA SciTech Forum, Orlando, FL,
  2024, {AIAA} 2024-2436.

\bibitem{nguyen2023solar}
L.~Nguyen, K.~Medina, Z.~McConnel, M.~S. Lake, {Solar Cruiser TRAC} boom
  development, in: AIAA SciTech Forum, National Harbor, MD, 2023, {AIAA}
  2023-1507.

\bibitem{diedrich2023solar}
B.~Diedrich, Solar sail torque model characterization for the {Near Earth
  Asteroid Scout} mission, in: 6th International Symposium on Space Sailing,
  New York, NY, 2023.

\end{thebibliography}

\end{document}